\documentclass{rQUF2e}

\usepackage{epstopdf}% To incorporate .eps illustrations using PDFLaTeX, etc.
\usepackage{subfigure}% Support for small, `sub' figures and tables
\usepackage{caption} 
\theoremstyle{plain}

\theoremstyle{definition}

\theoremstyle{remark}

\begin{document}

%\jvol{00} \jnum{00} \jyear{2014} \jmonth{October}

\title{Canonical sectors and evolution of firms in \\the US stock markets}

\author{Lorien X. Hayden, Ricky Chachra, Alexander A. Alemi, \\ Paul H. Ginsparg and James P. Sethna$^{\ast}$\thanks{$^\ast$Corresponding author. Email: sethna@lassp.cornell.edu}}

\affil{Department of Physics \\Cornell University, Ithaca, NY 14853 USA} \received{November 17th, 2016}

\maketitle

\section*{Contact Information}
\begin{itemize}
\item Lorien X. Hayden
	\begin{itemize}
	\item (573) 819-2155
	\item lxh3@cornell.edu
	\end{itemize}
\item Ricky Chachra
	\begin{itemize}
	\item (212) 729-7529
	\item rickychachra@gmail.com 
	\end{itemize}
\item Alexander A. Alemi
	\begin{itemize}
	\item (814) 422-5364
	\item alexalemi@gmail.com
	\end{itemize}
\item Paul H. Ginsparg
	\begin{itemize}
	\item (607) 255-7371
	\item ginsparg@cornell.edu
	\end{itemize}
\item James P. Sethna
	\begin{itemize}
	\item (607) 275-7748 
	\item sethna@lassp.cornell.edu
	\end{itemize}
\end{itemize}

\section*{Acknowledgements}
We thank Jean--Philippe Bouchaud, Ming Huang  and Janet Gao for helpful discussions.

\section*{Funding}
This work was partially supported by NSF grants DMR-1312160, DMR-1719490, IIS-1247696 and DGE-1144153.

\clearpage 

\section*{Abstract}
\begin{abstract}
A classification of companies into sectors of the economy is important for
macroeconomic analysis and for investments in sector-specific financial
indices and exchange traded funds (ETFs). Major industrial classification
systems and financial indices have historically been based on expert opinion
and developed manually. Here we show how unsupervised machine learning can
provide a more objective and comprehensive broad-level sector decomposition 
of  stocks. An emergent low-dimensional structure in the space of
historical stock price returns automatically identifies 'canonical sectors'
in the market, and assigns every stock a participation weight into these
sectors. Furthermore, by analyzing data from different periods, we show how
these weights for listed firms have evolved over time.
\end{abstract}

\begin{keywords}
Machine Learning, Archetypal Analysis, Canonical Sectors, Computational Finance
\end{keywords}

\begin{classcode}C38, G10\end{classcode}

\section*{Main Text}

Stock market performance is measured with aggregated quantities called indices
that represent a weighted average price of a basket of stocks. 
Market-wide indices such as Russell 3000\textsuperscript{\textregistered} \citep{Russell} and the S\&P 500\textsuperscript{\textregistered} \citep{sp500} consist of stocks from diverse companies reflecting a broad cross-section of the market. Sector-specific indices such as the Dow Jones\textsuperscript{\textregistered} Financials Index \citep{dowjones}, CBOE\textsuperscript{\textregistered} Oil Index \citep{cboe} and the Morgan Stanley\textsuperscript{\textregistered} High-Tech 35 Index \citep{mstech}, etc., are more granular and their composition requires a classification of companies into sectors. Major industrial classification schemes classify firms into sectors, albeit with many ambiguities \citep{Nadig11}. It is not clear, for example, how to assign a sector to conglomerates or diversified companies such as General Electric\textsuperscript{\textregistered}. Conversely, non-conglomerates with exposure to firms outside their own sector (for example, an investment bank exclusively serving pharmaceutical firms) also blur the boundaries of sector-identification. Moreover, as companies and their economic environments evolve, neither the industrial sectors nor the firms' sector association remains static, necessitating updates to sector assignments and addition of new sectors. 

\begin{figure}[h]
	\begin{centering}
	\includegraphics[width=0.4\textwidth]{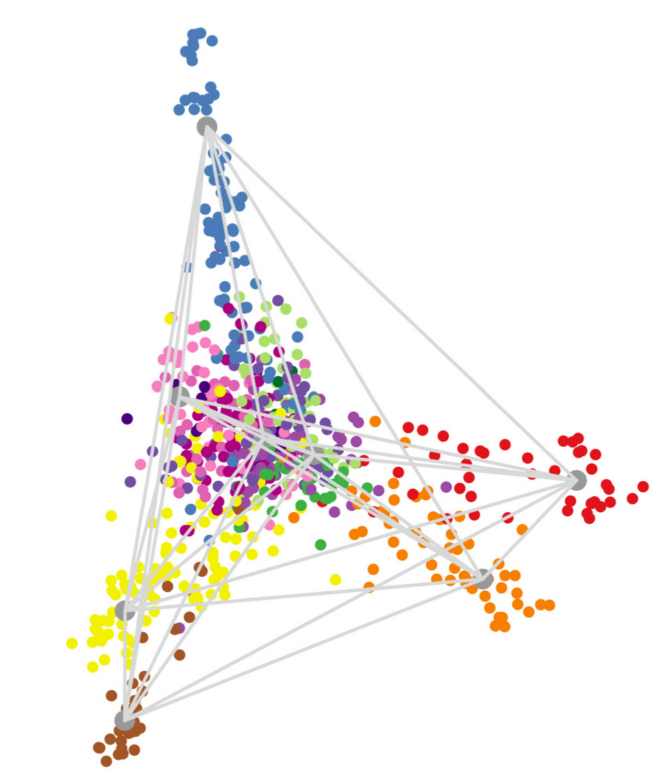}
	\caption{\label{fig:tetra01}
	\textbf{Low-dimensional projection of the stock price return data.} Stock
	price returns are projected onto a plane spanned by two stiff vectors from the
	SVD of the emergent simplex corners as described in Appendix~\ref{sec:projections}.
	Each colored circle corresponds to one of the 705 
	stocks in the dataset used in the analysis.  Colors denote the sectors assigned to 	
	companies by Scottrade\textsuperscript{\textregistered} \citep{scottrade} with the color scheme of Fig.~\ref{fig:Csf}.
	The grey corners of the simplex correspond to 
	sector-defining prototype stocks, whereas all other circles are given by a suitably
	weighted sum of these grey corners.  Projections along other singular vectors are 
	shown in Fig.~\ref{fig:raw2alltetra}.}
	\end{centering}
\end{figure}

\begin{figure}[h]
	\begin{centering}
	\includegraphics[width=0.5\textwidth]{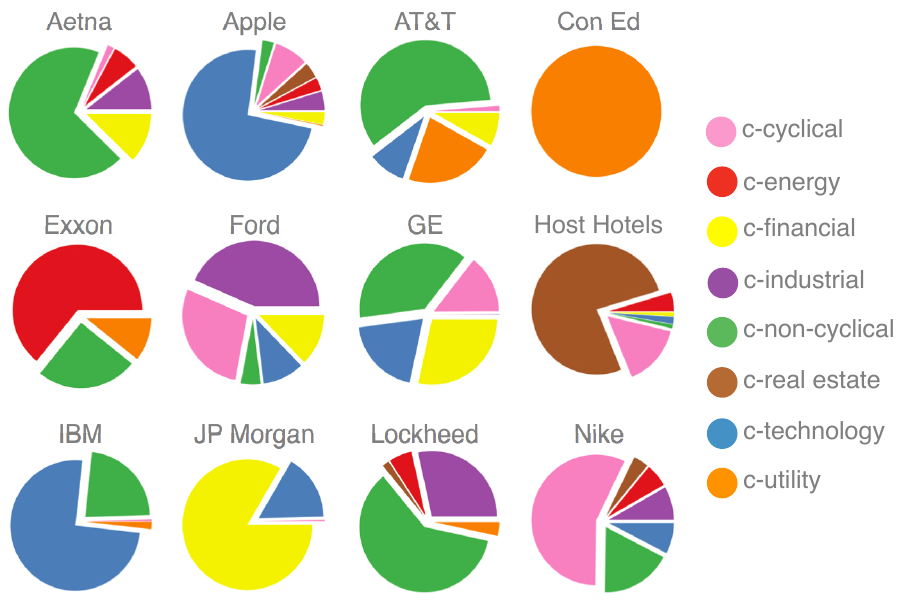}
	\caption{\label{fig:pies} 
	\textbf{Canonical sector decomposition} of stocks of selected companies. 
	A complete set of all 705 stocks is provided on the companion website \citep{site};
         the color scheme is shown on the right. Conglomerates like 
	GE\textsuperscript{\textregistered} decompose
         roughly into their core business lines. Tech firms such as 
	Apple\textsuperscript{\textregistered} that
         sell mass-market consumer goods have an important fraction in
         \textit{c-cyclical}, whereas IBM\textsuperscript{\textregistered} has a significant portion of
         \textit{c-non-cyclical} returns presumably due to its government
         contracts. Telecom companies like AT\&T\textsuperscript{\textregistered} are generally 	
	classified under  a separate telecom category by major classification systems, yet
         analysis shows their returns are described by a combination of
         \textit{c-non-cyclical} and \textit{c-utility} sectors. Health
         insurance providers like Aetna\textsuperscript{\textregistered} are commonly classified as 
	financial services firms, but their returns consist of a major part
         \textit{c-non-cyclical} and only a minor part of
	\textit{c-financial}---the healthcare sector is generally less prone to
         economic downturns. Defense contractors like Lockheed\textsuperscript{\textregistered} 
	are listed as capital goods companies, but their returns are seen to be majority
	\textit{c-non-cyclical} and only a smaller share of
	\textit{c-industrial} sector.} 
	\end{centering}
\end{figure}

A significant number of studies have previously aimed at identifying categories of stocks in financial markets with a variety of approaches. Recent numerical techniques have included extensive use of random matrix theory, principal component analysis or associated eigenvalue decomposition of the correlation matrix \citep{Plerou02,Kim05,Fenn11,Conlon09,Eom07,Coronnello05}, specialized clustering methods \citep{Mantegna99,Bonanno00,Bonanno03,Heimo09,Basalto05,Kullmann00,Musmeci14}
or time series analysis \citep{Podobnik08,Martins07}, pairwise coupling analysis \citep{Bury13}, and even topic-modeling of returns \citep{Doyle09}. Indeed, relevant prior work analyzing historical stock price returns \citep{Laloux99,Plerou02,famafrench} elucidated that the high-dimensional space of stock price returns has a low-dimensional representation. 

In parallel with this, there is a long tradition of style analysis in finance in which time series can be selected which serve as useful benchmarks for the performance of other stocks or indices. The three-factor model of Fama and French \citep{famafrench} is one such example. Recently, D. Vistocco and C. Conversano \citep{Vistocco09} proposed that Archetypal Analysis (AA) \citep{Cutler94} could provide these benchmark time series while also providing a way to plot this data in a meaningful way. In particular, they provide a triangular plot for Italian mutual funds and suggest parallel coordinate plots or asymmetric maps for higher dimensional representations. 
The positive decomposition of mutual funds into sectors using standard benchmarks (not derived using AA) was later studied by the same authors \citep{ConversanoV10}.

\begin{figure}[h]
	\begin{centering}
	\includegraphics[width=0.8\textwidth]{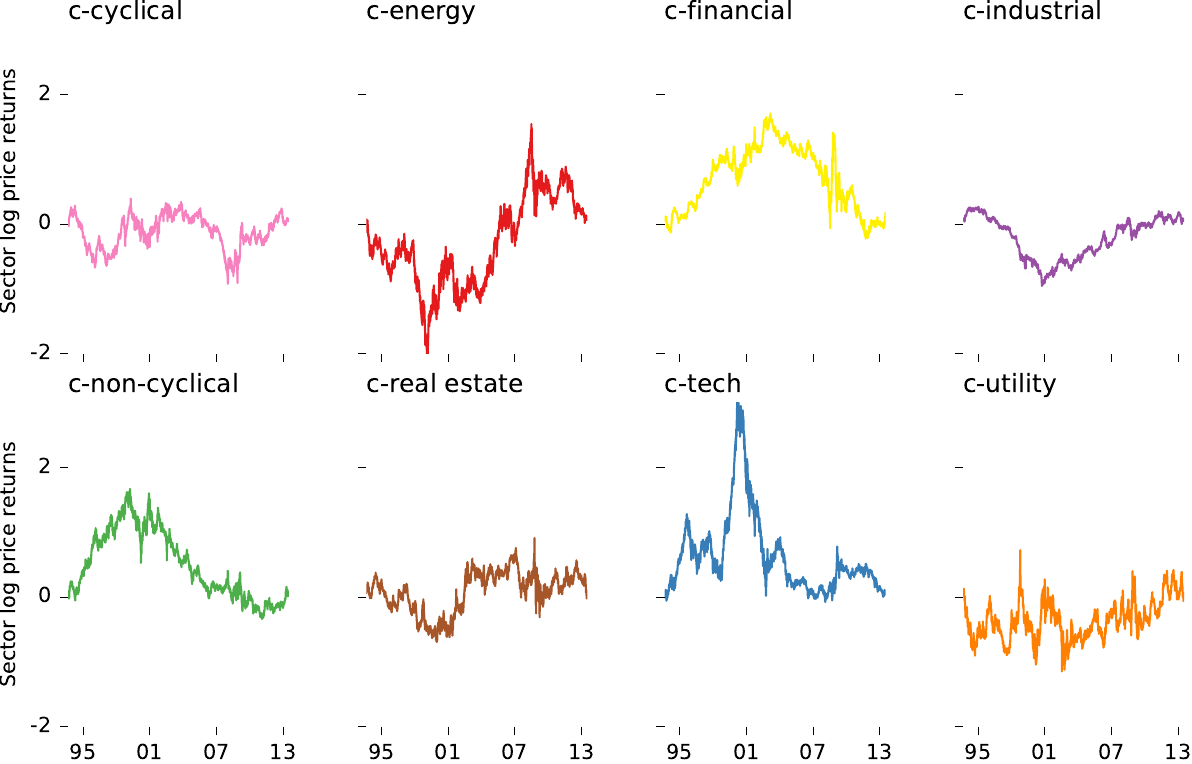}
	\caption{\label{fig:cumrets}
	\textbf{Emergent sector time series.} Annualized cumulative log price
	returns of the eight emergent sectors are shown. The time series capture
	all important features affecting different sectors: building-up of the dot-com
	bubble (c.\ 2000) followed by a burst, the soaring energy valuations (2003--08)
	followed by a crash, and the financial crisis of 2008.  We note that the dot-com
	bubble was confined to the c-tech sector whereas the financial crisis effects were
	spread throughout the sectors. Precise definition of the cumulative returns 
	plotted here is given in Eqn.~\ref{eq:Q}; other measures of sector dynamics are in Fig.~\ref{fig:retpanels}.} 
	\end{centering}
\end{figure}

\begin{figure}[h] 
	\begin{centering} 
	\includegraphics[width=0.8\textwidth]{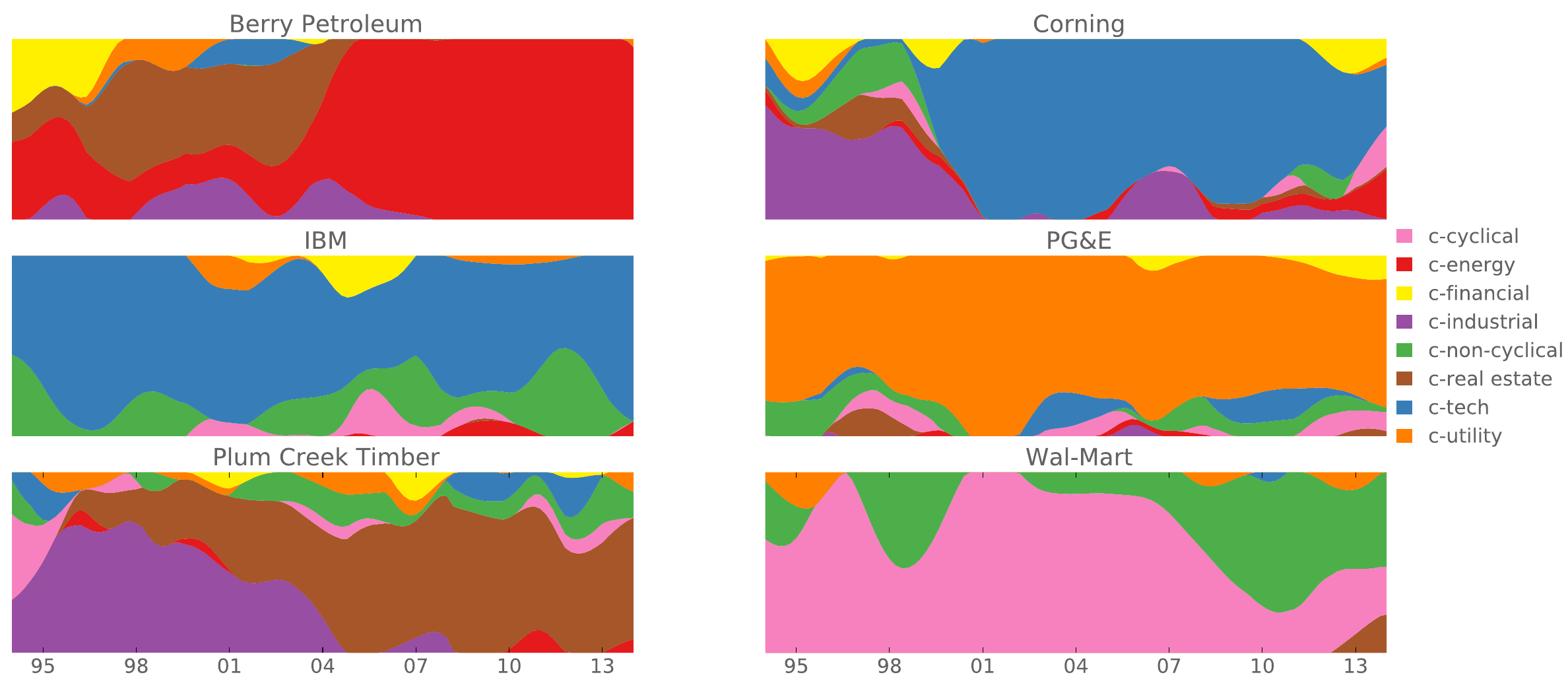} 
	\caption{\label{fig:flows}
	\textbf{Evolving sector participation weights.} Results from the sector
	decomposition made with rolling two-year Gaussian windows are shown for
	selected stocks. A complete set of 705 charts is provided on the companion
	website \citep{site}.
	% Color scheme is as in Fig.~\ref{fig:pies}. 
	For stable and
	focused companies such as Pacific Gas \& Electric\textsuperscript{\textregistered} or 
	IBM\textsuperscript{\textregistered}, one sees no
	significant shifts in sector weights; changes in time agree with errors expected from 
	unresolved fluctuations  \citep{site}. Wal-Mart\textsuperscript{\textregistered}'s returns, 
	on the other hand, 	have moved significantly from \textit{c-cyclical} to \textit{c-non-cyclicals} 
        (consumer staples) in the post-financial crisis years as shown; this is also
	true of other low-price consumer commodities retailers such as 
	Costco\textsuperscript{\textregistered}, but not true of higher price retailers such as 
	Whole Foods\textsuperscript{\textregistered}, Macy's\textsuperscript{\textregistered}, 
	etc. Corning\textsuperscript{\textregistered}, previously an \textit{industrial} firm with a 
	huge presence in optical fiber, suffered in the aftermath of the dot-com crisis 
	and now is classified as a \textit{tech} firm presumably due to
	its Gorilla\textsuperscript{\textregistered} glass used in cellphones,
	laptop displays, and tablets. Berry Petroleum grew within its
	home state of California in the early 1990s through development on properties
	that were purchased in the earlier part of 20th century. In 2003, the company
	embarked on a transformation \citep{berry} by direct acquisition of light oil
	and natural gas production facilities outside California. The figure shows a
	clear shift in the distribution of sector weights as the company has moved
	toward \textit{c-energy} and away from \textit{c-real estate}. Similarly, as
	Plum Creek\textsuperscript{\textregistered} Timber converted to a real estate investment 
	trust (REIT) in the late 1990s \citep{plumcreek}, its sector weights have significantly shifted
	toward \textit{c-real estate} sector.} 
	\end{centering}	
\end{figure}

Here, we demonstrate a new, holistic way of classifying stocks into industrial sectors by utilizing the emergent structure of price returns in data space.  Beyond the proposal of Vistocco and Conversano, we provide an interpretation of the archetypes of AA as sectors of the economy. This structure is purely contained in the geometry of the time series. Other methods, such as SVD, can discern that there is some such structure but are not well suited to a clean description. Archetypal Analysis, on the other hand, determines the convex hull of the dataset making it uniquely suited to creating a quantitative analysis of  the data.  In particular, if we take the log price returns of individual stocks, remove the overall market return, normalize to zero mean and unit s.d., then stock returns are well-approximated by a hyper-tetrahedral structure. Each lobe of the hyper-tetrahedron is populated by stocks of similar or related businesses (Fig.~\ref{fig:tetra01}); the lobe-corners (\textit{canonical sectors}) approximate the returns of companies that are prototypical of individual sectors (Table~\ref{table:constituents}). Returns of each stock can be decomposed into a weighted sum (Fig.~\ref{fig:pies}) of the canonical sector returns (Fig.~\ref{fig:cumrets}). Lastly, the canonical sector weights for a given company are dynamic and lead to insights into its evolution (Fig.~\ref{fig:flows}).

\begin{figure*} [h] %This figure requires two columns
	\begin{centering}
	\includegraphics[width=0.8\textwidth]{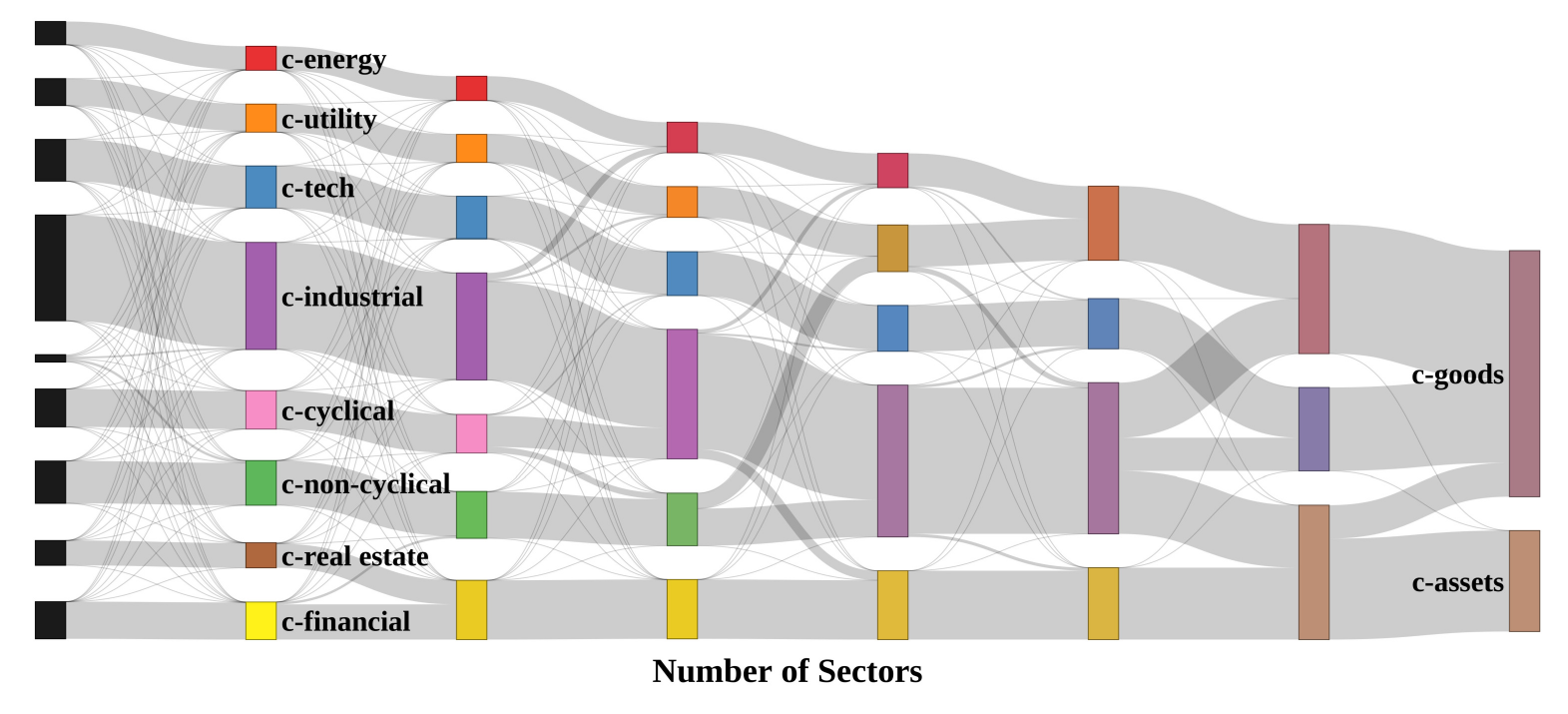}
	\caption{\label{fig:sankey}
	\textbf{Changes in the decomposition with dimensionality}. A Sankey diagram 
	(generated using D3 \citep{D3}) displaying the relationships between sector decompositions 
	with $n=N+1$ and $n=N$. Relative node sizes correspond roughly to the amount of the 
	market participating in the sector. Connection width depicts how strongly the sectors for 
	decompositions with different $n$ relate. For details, see Appendix
	section~\ref{subsec:DimChanges}. }
	\end{centering}
\end{figure*}

\begin{table*}
\begin{center}
\begin{minipage}{0.9\textwidth}
\tbl{\textbf{Canonical sectors and major business lines of primary
         constituent firms.} The eight canonical sectors identified by the
   	analysis described here are listed in the column on the left; these were
	named in accord with the business lines (middle column) of firms that show
	strong association with these sectors. Some examples are provided in the right
	column; a full list is available on companion website \citep{site}.}
{\begin{tabular}{|c|c|c|}
		\hline
         		\textbf{Canonical sector} & \textbf{Business lines} & \textbf{Prototypical 	
		examples} \\ 
		\hline 
		\textit{c-cyclical} & general  and specialty retail, discretionary goods & 		
		Gap, Macy's, Target \\
		\textit{c-energy} & oil and gas services, equipment, operations & 
		Halliburton, Schlumberger\\ 
		\textit{c-financial} & banks, insurance (except health) & 
		US Bancorp., Bank of America \\ 
		\textit{c-industrial}  & capital goods, basic materials, transport &
		Kennametal, Regal--Beloit \\ 
		\textit{c-non-cyclical} & consumer staples, healthcare &  
		Pepsi, Procter \& Gamble \\ 
		\textit{c-real estate} & realty investments and operations & Post Properties, 
		Duke Realty \\
		\textit{c-technology} & semiconductors, computers, comm.\ devices &
     		Cisco, Texas Instruments \\ 
		\textit{c-utility} & electric and gas suppliers & 
		Duke Energy,	Wisconsin Energy\\  	
		\hline 
	\end{tabular}}
\label{table:constituents} 
\end{minipage}
\end{center}
\end{table*}

The matrix of daily log returns of a stock $s$ are defined as
$r_{ts}=\log P_{ts} - \log P_{(t-1)s}$ where $P_{ts}$ are adjusted
closing prices (\textit{i.e.} corrected for stock splits and dividend
issues) and $t$ is in trading days. In the present analysis, we used
normalized returns, $R'_{ts}= (r_{ts} - \langle r_{ts}\rangle_t)/
\sigma_s$, where $\sigma_s^2= \langle r_{ts}^2 \rangle_t-\langle
r_{ts}\rangle_t^2$ is the variance (squared volatility) and $\langle \rangle_t$
represents the average over time (trading days). Overall market
returns from each stock were also removed, 
yielding what we shall call the log price returns
$R_{ts}=R'_{ts}-\langle R'_{ts}\rangle_s$.
(The two degrees of freedom we remove from each stock -- the variance and the overall return -- are of practical interest elsewhere, but obscure the
classification into sectors.)
The
hyper-tetrahedron, or simplex, which emerges (Fig.~\ref{fig:tetra01}) is
a self-organized structure: it has prototypical firms in corners
(Table~\ref{table:constituents}), closely related firms clumped together
in each lobe, diversified companies
(GE\textsuperscript{\textregistered}, Walt
Disney\textsuperscript{\textregistered},
3M\textsuperscript{\textregistered}, etc.) close to the center, and the
number of lobes denoting how many distinct sectors are exhibited by the
data.  This suggests a natural way to decompose stocks into canonical
sectors: for convex sets, each interior point is representable as a
unique weighted sum of corner points, implying here that every stock's
return is approximated by a weighted sum of returns from the
canonical sectors. Conversely, the weights for a given stock quantify
its exposure to the canonical sectors.

We applied an in house python implementation of the AA algorithm described by M\o rup and Hansen \citep{Morup12}.  The dataset consisted of 705 US firms' stocks with a minimum \$1 billion June 2013 market capitalization and with continuous 20 years (1993--2013) of listing on major exchanges (Appendix~\ref{sec:dataset}). Analysis of this dataset (Appendices~\ref{sec:factorization} and~\ref{sec:calculations}) revealed eight emergent sectors which were named in accordance with the companies they comprised (prefix \textit{c-} denotes ``canonical''): \textit{c-cyclical} (including retail), \textit{c-energy} (including oil and gas), \textit{c-industrial} (including capital goods and basic materials), \textit{c-financial}, \textit{c-non-cyclical} (including healthcare and consumer non-cyclical goods), \textit{c-real estate}, \textit{c-technology}, and \textit{c-utility}.  Calculated participation weights for a sample of 12 firms in Fig.~\ref{fig:pies} show a decomposition of their stocks into the canonical sectors with resulting insights discussed in the caption. Associated with each canonical sector $f$ is a time series of returns. As expected, these series show hallmark historical events of individual sectors (Fig.~\ref{fig:cumrets}): the dot-com bubble, the energy crisis, and the financial crisis being the major events in the last two decades.  

Determining the correct number of canonical sectors that appropriately describe the space of stock market returns is akin to the more general issue of selecting a signal-to-noise ratio cutoff, or a truncation threshold in the dimensional-reduction of data. The choice of this threshold is generally sensitive to sampling, yet the results presented here are reasonably robust with different choices leading to meaningful and similar decompositions. Fig.~\ref{fig:sankey} depicts the changes in the decomposition with dimension. Details of how the figure was generated as well as more information on the two and three dimensional decompositions are available in Appendix~\ref{sec:numbern}.

 In addition to the full data set of 20 years $\times$ 705 firms, we also  applied the algorithm to overlapping, two-year Gaussian windows to study how the sector weights for firms have evolved in time (Fig.~\ref{fig:flows}, see
also
Appendix~\ref{sec:calculations}).  As expected, the sector decomposition of firms is dynamic. Mergers, acquisitions, spin-offs, new products, effect of competitive environments or shifting consumer preferences can change the business foci of firms and hence alter the sector association of firms. External events affecting companies in an idiosyncratic manner also show clear signature in this analysis. 

The eight-factor decomposition presented here explains 11.1\% of the total  variation ($r^2$) in the normalized returns with the market mode removed, and 56\% of the random matrix theory explainable variation  defined in Appendix~\ref{sec:pve}.  For comparison, the classic three-factor decomposition of portfolio returns by Fama and French \citep{famafrench} into market mode, market capitalization, and growth versus value  yields an $r^2$ value of only 4.75\%. Indeed, if only three factors are used instead of  the eight for the decomposition presented here, the regression yields a comparable $r^2$ value (5.61\%) but there appears to be no correspondence between three factors found by our unsupervised model, and those of Fama and French (Fig.~\ref{fig:comp}). Carrying out a similar comparison with Fama and French's analysis applied to model portfolio returns, the regression on the S\&P 500\textsuperscript{\textregistered} yields an $r^2$ value of 99.4\% for Fama and French compared to 93.5\% for our eight-factor decomposition (market mode reintroduced). Our decomposition was optimized without concern for market capitalization, which appears to be the key difference: For an equal weighted index of the 338 stocks in the S\&P 500\textsuperscript{\textregistered} with current tickers and a complete data series in our time of interest, we obtain an $r^2$ value of 99.0\% (97.0\% for 3 factors) compared to 95.8\% for Fama and French.  
We conclude that a sector decomposition like the one presented here, perhaps weighted by market
capitalization, should be an improved guide to investors, compared to the widespread value/growth
and large-cap/small-cap stock characterizations currently used.

Future work remains to address survivorship bias, effects of sampling at different frequencies, and incorporating market capitalization.  Investors, analysts, and governments alike would benefit from the development of new investable sector indices (Appendix~\ref{sec:indices}) that measure the health of our industrial sectors just like the macroeconomic indicators (GDP, housing starts, unemployment rate, etc.) measure the health of our broader economy. Tracing the sectors back in time could elucidate the incorporation of science and technology into our economic system. Finally, our unsupervised decomposition could provide data suitable for quantitative modeling of the internal and external dynamics of our economic system. 

%\section*{Supplemental material}
%The supplement contains analysis complementary but non-essential to the main text. These include dataset particulars, details of the algorithm convergence, and discussion on the coefficient of determination, sector changes with dimensionality and robustness

\appendix

\section{Dataset Particulars}
\label{sec:dataset}

Company names, tickers, listed-sectors and market caps of US-based firms used
in this analysis were obtained from Scottrade\textsuperscript{\textregistered} \citep{scottrade}. Daily closing prices adjusted for stock splits and dividend issues were obtained from Yahoo\textsuperscript{\textregistered} Finance \citep{yahoo}. The rare cases of missing prices in the time series were replaced with linearly interpolated values. A brief summary of listed sectors and number of companies in each is provided in Table~\ref{tab:listed} and a full list of company names, tickers, market caps and listed-sector info is
available on the companion website \citep{site}.

\begin{center}
	\begin{table}[h]
		\centering
		{\footnotesize
		\begin{tabular}{c | c}
			\textbf{Listed sector} & \textbf{Companies} \\
			\hline 
			Basic materials & 58  \\
			Capital goods & 61 \\ 
			Consumer cyclical & 41 \\
			Consumer non-cyclical & 40 \\
			Energy & 42 \\
			Financial (+Real estate) & 138 \\
			Healthcare & 53 \\
			Services (+Retail) & 101 \\
			Technology & 93 \\
			Telecom & 6 \\
			Utility & 57 \\
			Transport & 15 \\
			\hline
			TOTAL & \textbf{705} \\
			 \hline
		\end{tabular}}
		\caption{\textbf{Listed sectors and number of companies dataset analyzed.} Tickers 	
		for each company were obtained from \citep{scottrade}.}
		\label{tab:listed}
	\end{table}
\end{center}

\section{Returns Factorization and Sector Decomposition}
\label{sec:factorization}

\begin{figure*}[h] 
	\begin{centering} 
	\includegraphics[scale=0.6]
	{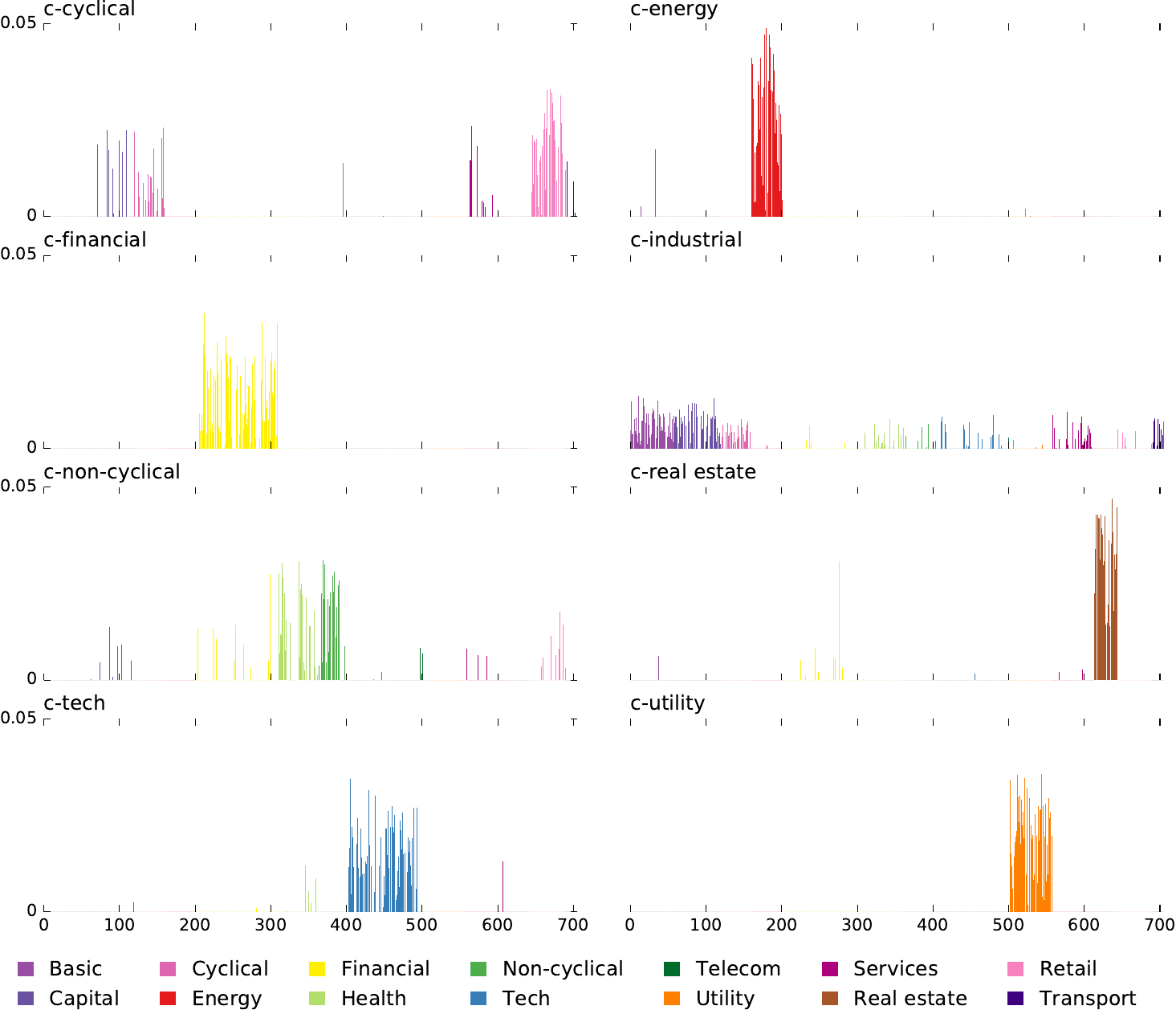} 
	\caption{\label{fig:Csf}
	\textbf{Canonical Sector Constituents} (shown as columns of the $C_{sf}$).
	$C_{sf}$ represents a weighted combination of stocks that defines the canonical
	sector each of which has a time series represented by $E_{tf}$ that is given by
	$E_{tf}=R_{ts}C_{sf}$. The eight subplots show the constituent participation
	component of stocks in each canonical sector $f$. Canonical sectors are labeled
	on the plot; their names were chosen according to the listed sectors of firms
	that comprise them. Noteworthy features seen above include the co-association
	of listed sectors: basic, capital, transport and part of cyclicals into
	\textit{industrial goods}. Similarly, healthcare and non-cyclicals are coupled
	together in what we call \textit{non-cyclicals}. Canonical \textit{retail} goes
	primarily with listed retail and cyclicals. Stocks are colored by listed
	sectors as shown at the bottom. Listed sector information was obtained
	from \citep{scottrade}.}
	\end{centering}
\end{figure*}

A variety of factorization algorithms have been developed in recent years for
dimensional reduction, classification or clustering. Examples include
archetypal analysis (AA) \citep{Cutler94}, heteroscedastic matrix
factorization \citep{Tsalmantza12}, binary matrix factorization \citep{Zhang07},
K-means clustering \citep{Ding04}, simplex volume maximization \citep{Thurau10},
independent component analysis \citep{Hyvarinen00}, non-negative matrix
factorization (NMF) \citep{Lee99,Wang13} and its variants such as the semi- and
convex-NMF \citep{Ding10}, convex hull NMF \citep{Thurau11} and hierarchical
convex NMF \citep{Kersting10}, among others. Each method has a unique
interpretation \citep{Li06} and therefore, a successful application of any of
these methods is contingent upon the underlying structure of the data. 

The hyper-tetrahedral structure of log price returns seen in our analysis
motivates a decomposition so that each stock's return is a weighted mixture of
canonical sectors, constrained to lie in the convex hull of the data. Hence we
employ AA factorization which is defined as:

\begin{equation}
	\begin{matrix}
	R_{ts}  \sim R_{ts'} C_{s'f} W_{fs} \\
	C_{s'f} \geq 0, \sum_{s'}C_{s'f}=1,\\
	W_{fs}\geq 0,  \sum_f W_{fs}= 1.  
	\end{matrix}
	\label{eq:AA}
\end{equation}

Columns of $R_{ts}C_{sf} = E_{tf}$ are the emergent sector time series (basis
vectors) representing the $n$ corners of the hyper-tetrahedron, and $W_{fs}$
are the participation weights ($W_{fs}\geq0$) in sector $f$ so that $\sum_f
W_{fs} = 1$ for each stock $s$. The sector matrix $E_{tf}$ is within the convex
hull ($C>0$, $\sum_s C_{sf} = 1$) of the data $R_{ts}$. It can be found by
either minimizing the squared error with convex constraints in factorization as
originally proposed \citep{Cutler94}, or by making a convex hull of the dataset
and choosing one or more of its vertices to be basis vectors, or by making a
convex hull in low-dimensions and choosing one or more of its vertices to be
basis vectors \citep{Thurau09}, or by minimizing after initializing with
candidate archetypes that are  guaranteed to lie in the minimal convex set of
the data \citep{Morup12}.  The columns of the $C$ matrix are shown in
Fig.~\ref{fig:Csf}.

\section{Calculations and Convergence}
\label{sec:calculations}

\begin{figure*}[h] 
	\begin{centering} 
	\includegraphics[scale=0.85]
	{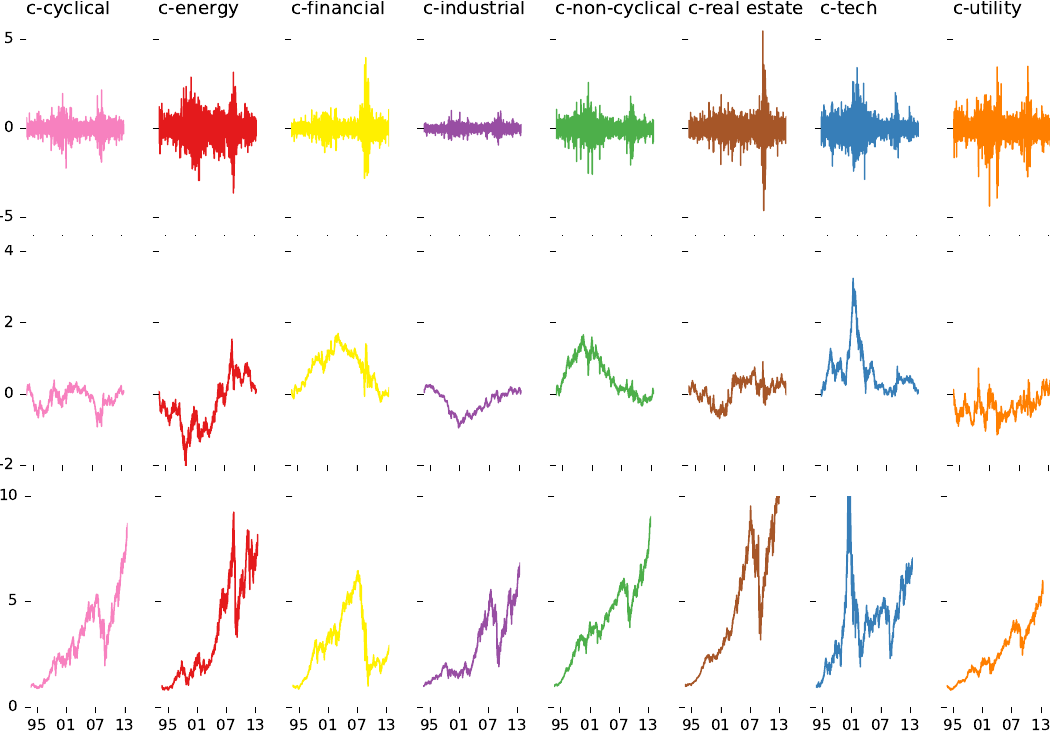}
	\caption{\label{fig:retpanels} 
	\textbf{Canonical sector time series.} Top
	row: normalized log returns (columns of $E_{tf}$), middle row: cumulative
	log returns (same as Fig.~\ref{fig:cumrets} and defined in
	Eqn.~\ref{eq:Q}), and bottom row: unweighted price index of canonical
	sectors (Eqn.~\ref{eq:index}).} 
	\end{centering}	
\end{figure*}

\begin{figure*}[h] 
	\begin{centering} 
	\includegraphics[width=0.85\textwidth]
	{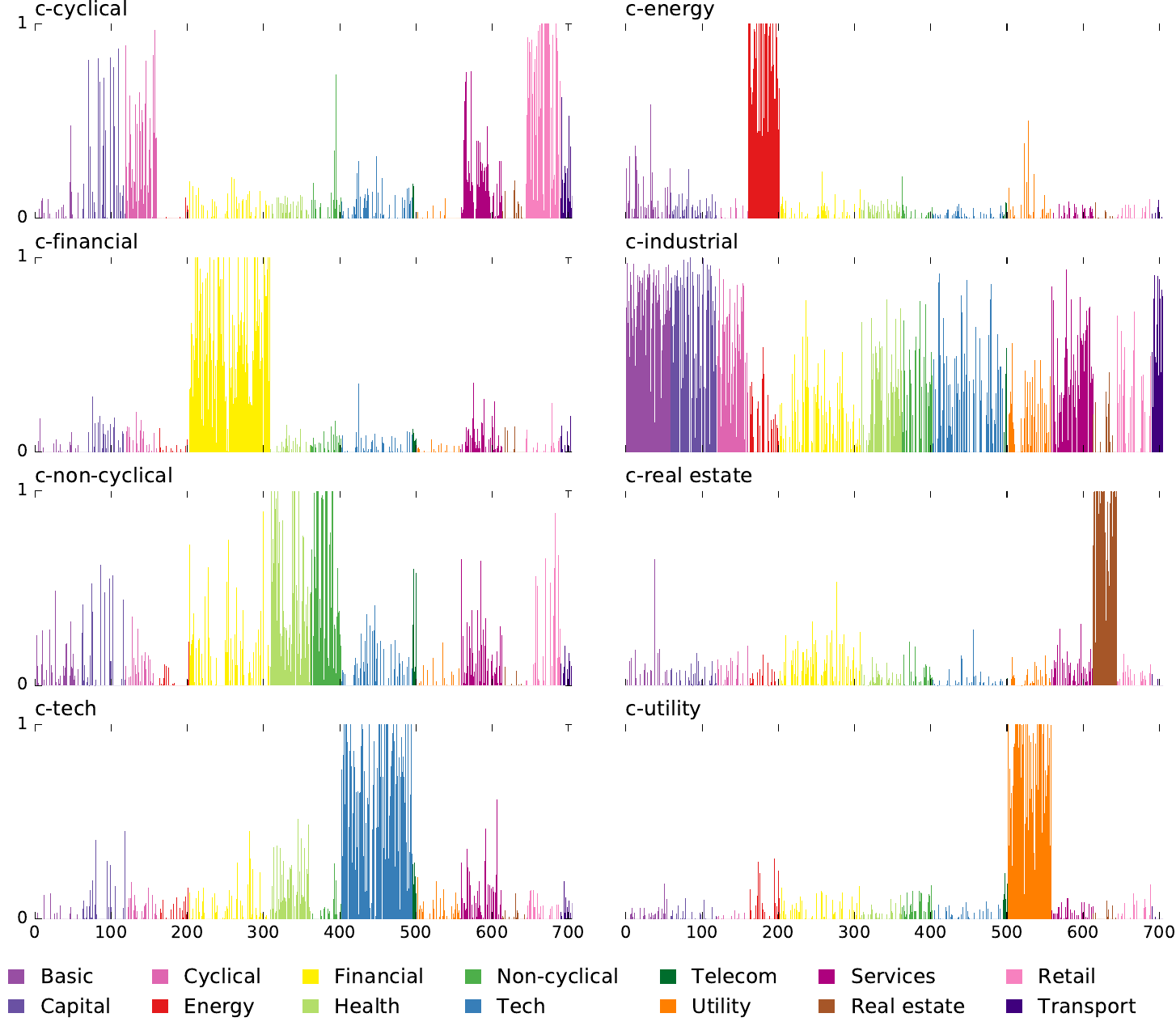} 
	\caption{\label{fig:Wfs}
	\textbf{Weight distribution in canonical sectors.} Each of the eight
	subplots shows the constituent participation weights of all 705 companies in a
	canonical sector (rows of $W_{fs}$). Stocks are colored by listed sectors as
	shown at the bottom. Listed sector information was obtained
	from \citep{scottrade}.} 
	\end{centering}
\end{figure*}

Numerical computations were performed using an in-house Python language
implementation of the principal convex hull analysis (PCHA) algorithm as
described in \citep{Morup12}. For the full dataset, the factorization $R=EW$,
with $E=RC$ as defined in Eqn.~\ref{eq:AA} converged in 35 iterations to a
predefined tolerance value of $\Delta_{SSE}<10^{-7}$, where $\Delta_{SSE}$ is
the average difference in the sum of squared error per matrix element in $R-EW$ from
one iteration to the next. The resulting columns of $E_{tf}$ are shown in Fig.~\ref{fig:retpanels} (top row). Annualized cumulative log returns are obtained
by summing rows of $E_{tf}$: 

\begin{equation} 
	Q_f(\tau) =\frac{1}{\sqrt{250}}\sum\limits_{t=0}^{t=\tau} E_{tf}
	\label{eq:Q}
\end{equation}

The time series $Q_f(\tau)$ are shown in Fig.~\ref{fig:cumrets} and the middle row of 
Fig.~\ref{fig:retpanels}. Weights $W_{fs}$ for selected stocks are shown in Fig.~\ref{fig:pies}, the remainder are available on the companion website \citep{site}. In each canonical sector $f$, the component of weights for companies are shown in Fig.~\ref{fig:Wfs}.

The analysis of evolving sector weights was performed similarly, but with a
sliding Gaussian time window. We  decomposed the local normalized log returns 
for each stock into the canonical sectors determined from the entire time
series. Each column (time series) of the returns matrix $R_{ts}$ was multiplied
with a Gaussian, $G_{\mu}(\tau)=\exp(-(\tau-\mu)^2/(2\times 250^2))$ of
standard deviation $250$ centered at $\mu$ to obtain $R^\mu_{ts}$. We use
$C_{s'f}$ found using the full dataset (Eqn.~\ref{eq:AA}) (corresponding to keeping
the sector-defining simplex corners fixed). $R^\mu_{ts}$
is factorized to obtain new weights $W^\mu_{fs}$ that describe sector
decomposition of stocks in that period focused at $t=\mu$:
$R^\mu=R^\mu_{ts'}C_{s'f}W^\mu_{fs}$. $\mu$ is increased in steps of 50
starting at $\mu=0$ and ending at $\mu=5000$, and $W^\mu$ is calculated at each
$\mu$ with the corresponding $R^\mu$. These results are plotted in
Fig.~\ref{fig:flows} for a select group of companies; the remainder are
available on the companion website \citep{site}.

\begin{figure*}[!ht]
	\begin{centering}
	\includegraphics[scale=0.7]
	{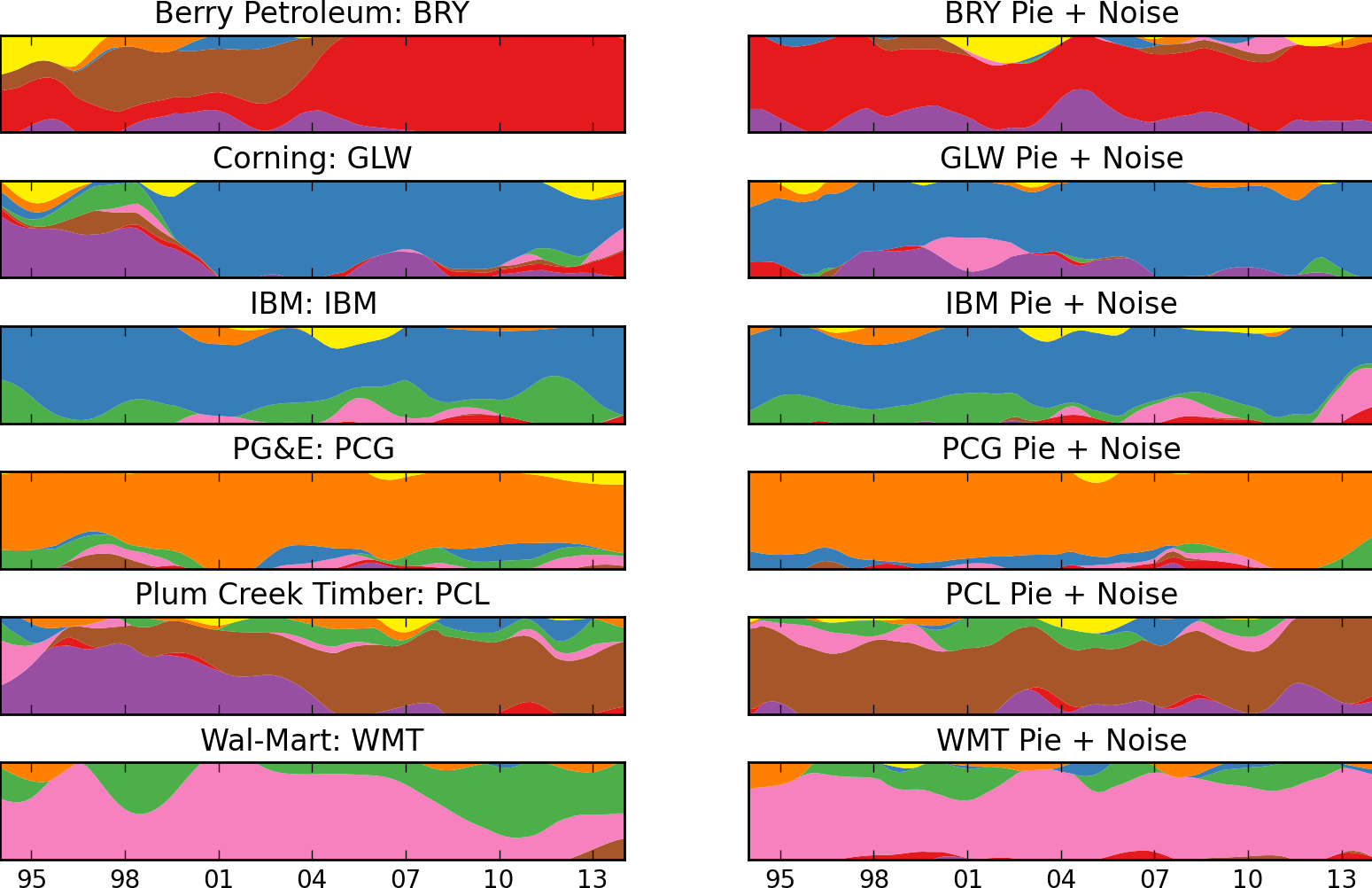}
	\caption{\label{fig:noisyflows}Comparison between flow diagrams presented in Fig.~\ref{fig:flows} with simulated data. The simulated data is created from the dot product of the weight vector of the company with the corner time series as described in this section.
This yields a version of the company with constant weights in time. To this we add gaussian noise with standard deviation one and repeat the analysis to generate the flows in time. In the left column are the actual flows for companies, on the right is their constant in time counterpart with added noise. We see that key features are in fact signal while small fluctuations correspond to noise. Color scheme as in Fig.~\ref{fig:pies} and Fig.~\ref{fig:flows}. }
	\end{centering}
\end{figure*}

To address the challenge of distinguishing signal from noise in the evolving sector weights,  we emulated the effect of noise for each of the companies from Fig.~\ref{fig:flows}.
%simulate data to which we add noise and then compare. This was done by repeating the analysis for the flows where the companies from Fig.~\ref{fig:flows} were replaced. 
For each of these companies, we took its sector weights, $\vec{\omega_f}$, and multiplied by $E_{tf}$ to obtain a time series for the company with weights that are constant in time. We then added gaussian random noise with standard deviation one and replaced these companies by this simulated data. Fig.~\ref{fig:noisyflows} shows the comparison between the real flows and the simulated constant data with noise added. General features are shown to be signal while small fluctuations are consistent with noise.

\section{Dimensionality of the Space of Price Returns}
\label{sec:dimensionality}

\begin{figure*}[h] 
	\begin{centering} 
% XXX	\includegraphics[scale=.75, trim= 10mm 15mm 20mm 15mm, clip, angle=270]
%	\includegraphics[scale=.55, trim= 10mm 15mm 20mm 15mm, clip, angle=270]
	\includegraphics[width=0.85\textwidth]
	{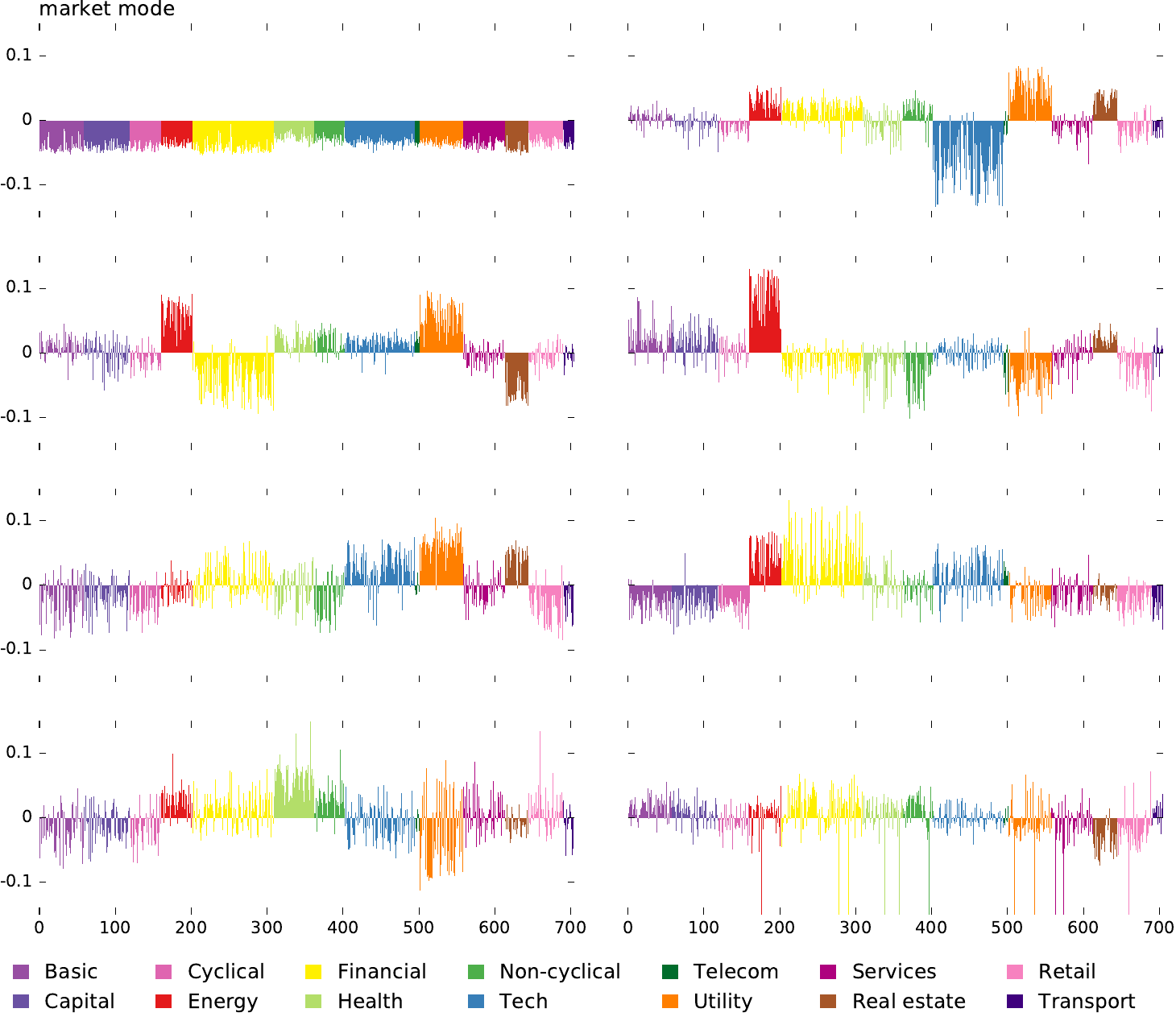} 
	\caption{\label{fig:Vtfs}
	\textbf{Singular vectors $V^T_{fs}$ of the SVD of returns $R_{ts}$.} The
	orthonormal right singular vectors (rows of $V^T_{fs}$) of SVD of $R_{ts}$ are
	equivalent to the eigenvectors of the stock-stock correlation matrix
	$\xi_{ss'}\sim R^TR$. Eight of these stiffest eigenvectors including the
	\textit{market mode} are shown in rows of two at a time. Each has 705
	components corresponding to stocks in  the dataset. The \textit{market mode}
	with all components in the same direction describes overall fluctuations in the
	market; it was excluded from the analysis described in the paper. Previous
	work \citep{Plerou02} has suggested that each eigenvector of the stock-stock
	correlation matrix describes a listed sector, however as seen above, a more
	correct interpretation is that each eigenvector is a mixture of listed sectors
	with opposite signs in components. For example, the stiffest direction (after
	market mode) has positive components in real estate and utility, but negative
	in tech. Less stiff eigenvectors (including the last one shown here), do not
	contain sector-relevant information. Stocks are colored by listed sectors as
	shown at the bottom. Listed sector information was obtained
	from \citep{scottrade}.} 
	\end{centering}
\end{figure*}

It is often the case with large datasets that the effective dimensionality of
the data space is much lower when one filters out the noise. Of the many
dimensional reduction methods, the most commonly used is singular value
decomposition (SVD) \citep{Press}, a deterministic matrix factorization. We
discuss SVD in more detail in order to draw a contrast with previous
SVD results, and to apply it for quantifying the explainable variation
in the returns data.

An SVD of $R_{ts}$ is a matrix factorization \citep{Press}
$R_{ts}=U_{tf}\Sigma_{ff'} V^T_{f's}$ such that matrices $U$ and $V$ are
orthogonal; $\Sigma$ is a diagonal matrix of ``singular values''. If the goal
were purely rank-reduction, $n$ entries of $\Sigma$ chosen to lie above ``noise
threshold'' are retained and the rest truncated so that $0\leq f,\ f'\leq n$.
This effectively reduces the dimension of $R$ to $n$. The choice of $n$ can be
informed by the distribution of singular values as discussed later. The rows of
$V^T$ are precisely the eigenvectors of the stock-stock returns correlation
matrix, $\xi_{ss'} \sim R_{st}^TR_{ts}$. It was previously reported that some
components of the stiff eigenvectors of this stock-stock correlation matrix
loosely corresponded to firms belonging to the same conventionally identified
business sector \citep{Plerou02} (but see Fig.~\ref{fig:Vtfs}).

\begin{figure}[h]
	 \begin{centering} 
% XXX	\includegraphics[scale=.80, trim= 10mm 10mm 115mm 130mm, clip, angle=270]
%	\includegraphics[scale=.60, trim= 10mm 10mm 115mm 130mm, clip, angle=270]
	\includegraphics[scale=.20]
	{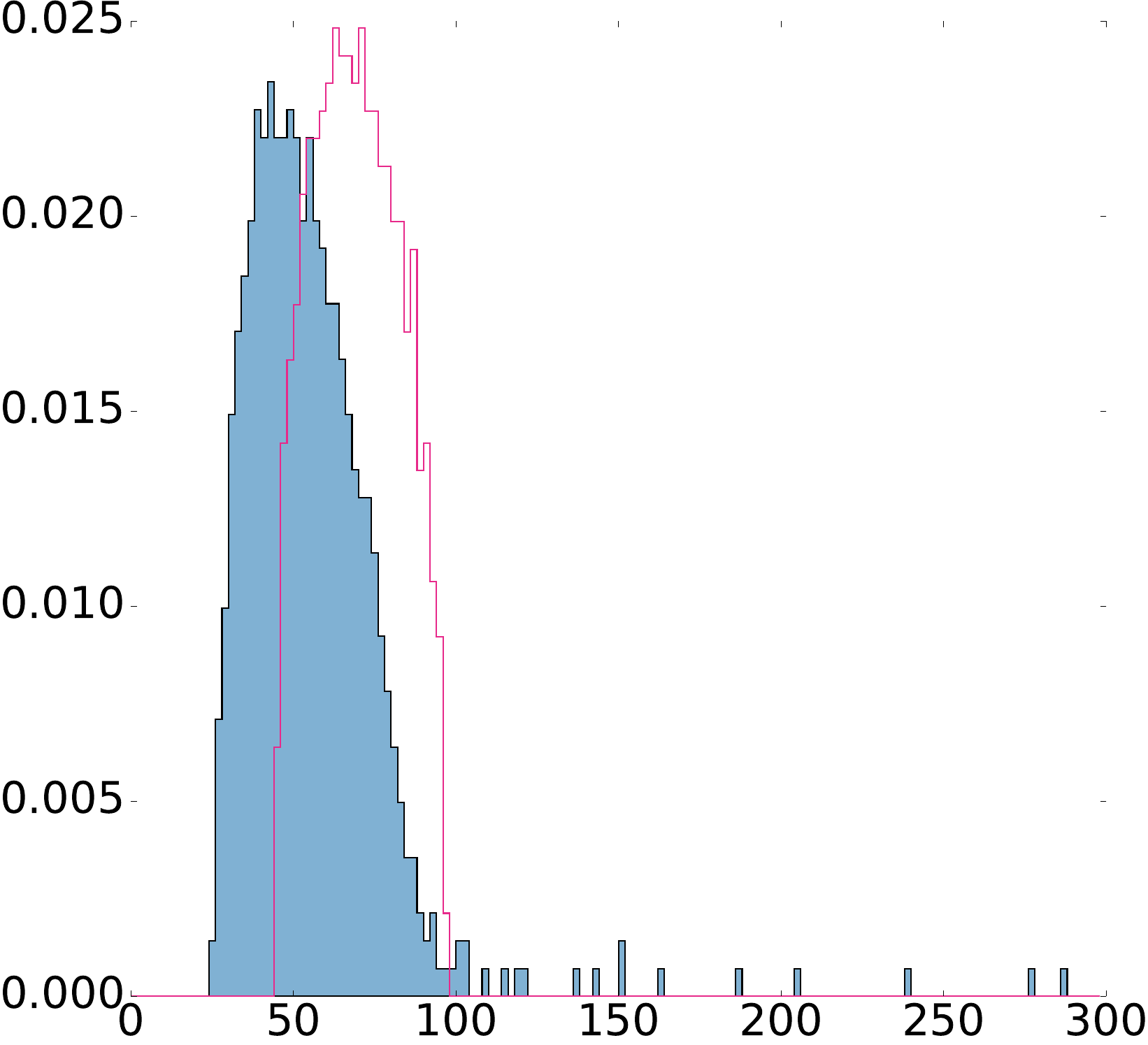}
	\caption{\label{fig:rmt_comp}
	 \textbf{Normalized distribution of singular
	values.} Filled blue histogram corresponds to distribution of singular
	values of returns from the dataset $R_{ts}$---one notices a clear separation of
	the hump-shaped bulk of singular values, and
	about 20 stiff singular values (the largest singular value $\sim$952,
	corresponding to the \textit{market mode} is not shown). Pink line histogram
	outline shows the distribution of singular values of a matrix of the same shape
	as $R$ but containing purely random Gaussian entries.} 
	\end{centering}
\end{figure}

After normalizing the log returns, the returns matrix $R$ has entries of unit
variance. If the entries were uncorrelated random variables drawn from a
standard normal distribution, their singular values (which are also the
positive square roots of the eigenvalues of $R^TR$) would be described by
Wishart statistics \citep{Mehta}. The Wishart ensemble for a matrix of size
$\alpha \times \beta$ predicts a distribution of singular values with a
characteristic shape \citep{Mehta}, bounded for large matrices by $\sqrt{\alpha}
\pm \sqrt{\beta}$. Comparing the stock correlations with Wishart statistics has
been previously used to filter noise from financial datasets \citep{Laloux99}. 
As shown in Fig.~\ref{fig:rmt_comp}, most singular values of the returns
matrix $R$ lie in the bulk below the bound set by the Wishart ensemble, whereas only $\sim$20 fall outside that cutoff (The singular value bounds of a random Gaussian rectangular
matrix of size $\alpha \times \beta$ can be shown to be
$\sqrt{\alpha}\pm\sqrt{\beta}$ for large  matrices.) Historically, this has served as indication that singular values within the bulk correspond to noise \citep{Laloux99}. Recently, however, much progress has been made in the development of techniques to extract signal from the bulk \citep{Burda04, Burda06, Livan11}. Our method does not claim to capture this information. Rather, we measure its ability to capture variation in the data above the cutoff by means of random matrix theory explainable variation as defined in 
section~\ref{sec:pve}. The largest singular value of $R_{ts}$ corresponds to what we will refer to as the ``market mode'' as this represents overall simultaneous rise and fall of stocks. In the analysis presented in this paper, this mode has been filtered from the returns
matrix by projecting the $R$ matrix into the subspace spanned by all non-market
mode eigenvectors. This is nearly equivalent to filtering the market mode using simple linear regression (as done commonly \citep{Plerou02}), although more convenient.  

\section{Low-Dimensional Projections of Price Returns}
\label{sec:projections}

\begin{figure*}[h] 
	\begin{centering} 
	\includegraphics[scale=0.4]
	{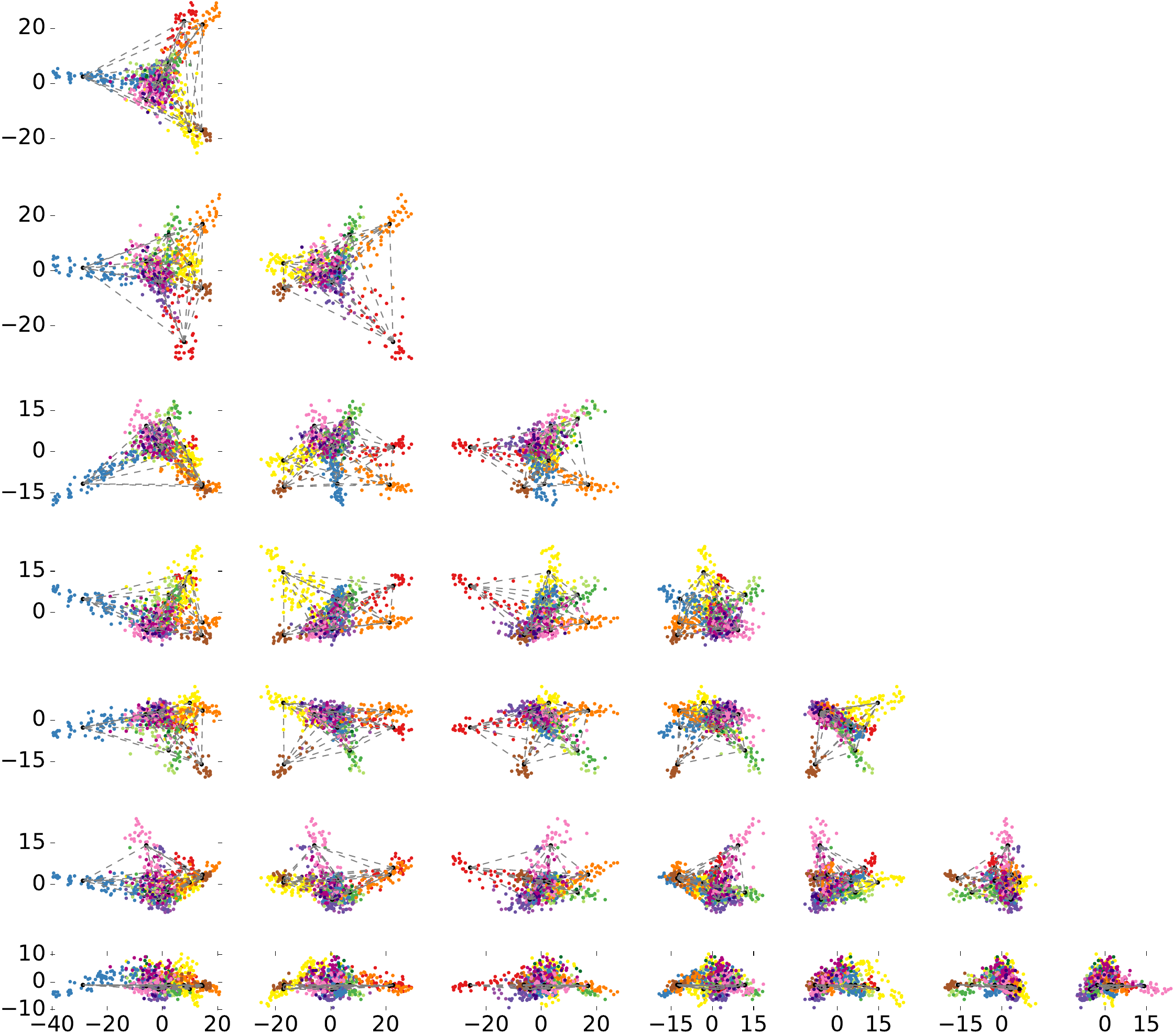}
	\caption{\label{fig:raw2alltetra} 
	\textbf{Low-dimensional projections of
	stock returns data}, colored by
	Scottrade\textsuperscript{\textregistered} sector.
	Each colored circle represents a stock in our dataset
	and is colored 
	according to sectors assigned by
        Scottrade\textsuperscript{\textregistered} \citep{scottrade}
	as indicated in Fig.~\ref{fig:Vtfs}. The first row is
	equivalent to Fig.~\ref{fig:tetra01}. Black circles represent
        the archetypes
	found with our analysis. The $(i,j)^{th}$ figure in the grid is a plane spanned
	by singular vectors $i$ and $j+1$ (rows of $X^TR$) from the calculations
	described earlier. Projections after the factorization are shown in
	Fig.~\ref{fig:alltetra}.} 
	\end{centering}	
\end{figure*}

\begin{figure*}[h] 
	\begin{centering} 
	% Old 
	%\includegraphics[scale=.88, trim= 15mm 13mm 15mm 75mm, clip, angle=270]
	%{Figures/alltet}
	% Ginsparg
	%\includegraphics[scale=.88]
	%{Figures/alltet.pdf}
	\includegraphics[scale=0.4]
	{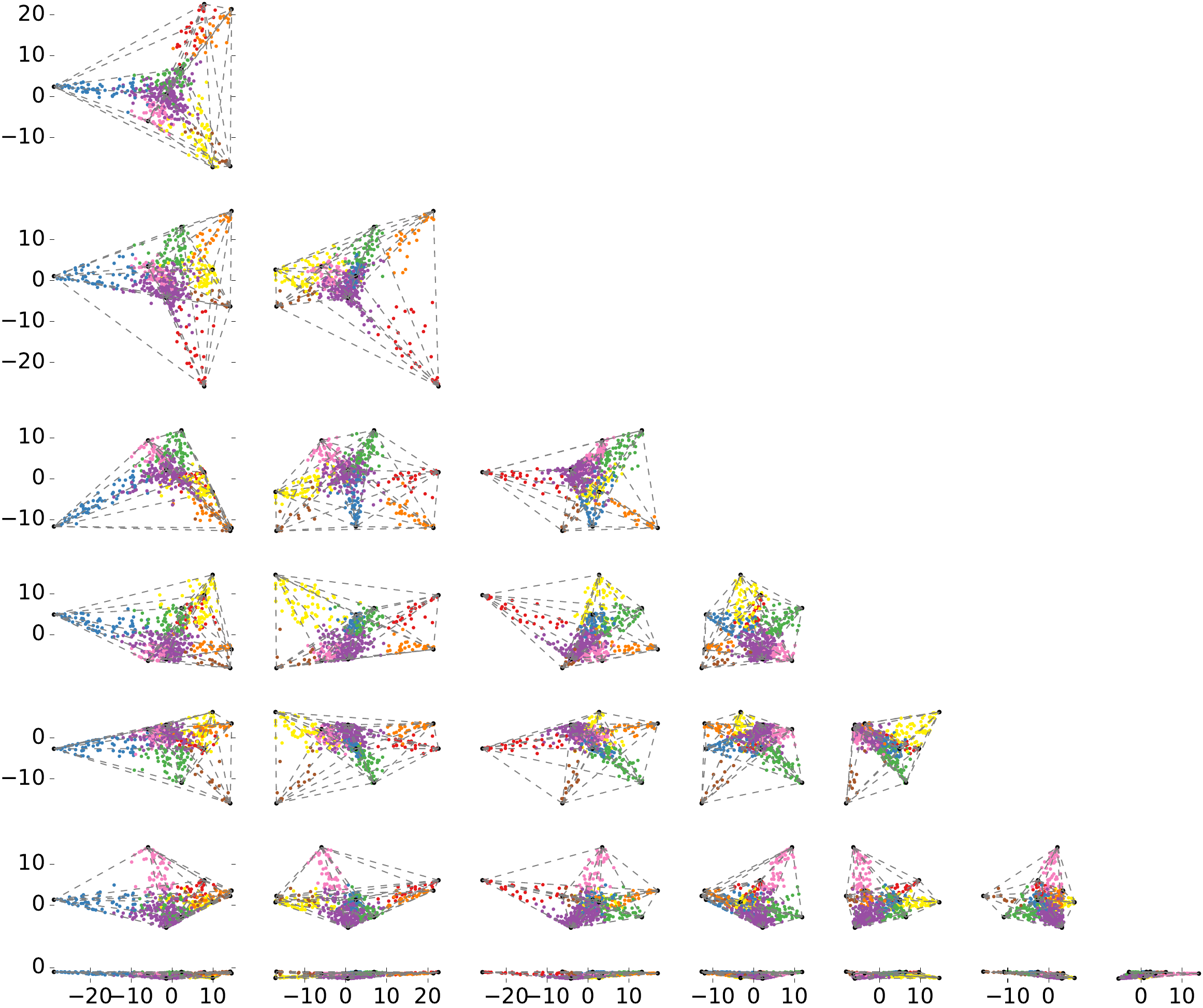}
	\caption{\label{fig:alltetra}
	\textbf{Cross-sections along eigenplanes of
	the factorized returns.} Each colored circle represents a stock in our
	dataset and is colored according to the primary canonical sector
	association with the color scheme in Fig.~\ref{fig:pies}.
	Black circles represent the archetypes found with our analysis. The
	$(i,j)^{th}$ figure in the grid is a plane spanned by singular vectors $i$ and
	$j+1$ (rows of $MN^T$) from the calculations described 
	earlier. Projections of raw data (before the factorization)
	are shown in Fig.~\ref{fig:raw2alltetra}. Note that the colors
	are very similar to those of the traditional
	Scottrade\textsuperscript{\textregistered} classification shown in
	Fig.~\ref{fig:raw2alltetra}; the color schemes were designed to roughly
	match. Note that here all points have been projected into the
	hyper-tetrahedron by our factorization.} 
	\end{centering}	
\end{figure*}

The emergent low-dimensional, hyper-tetrahedral (simplex) structure of stock
price returns can be seen by projecting the dataset into stiff ``eigenplanes''.
Eigenplanes are formed by pairs of right singular vectors from a SVD. Here,
we construct an SVD of the simplex corners, $E_{tf} = X_{tk}YZ^T_{kf}$; simplex
corners are mapped to columns of $YZ^T$ because $YZ_{kf}^T=X_{kt}^TE_{tf}$ (in
other words, $X^T_{kt}$ is a projection operator). The plots in
Fig.~\ref{fig:raw2alltetra} are the projections of the dataset,
$X^T_{kt}R_{ts}=v_{ks}$. The rows of $v$ taken in pairs form the axes of the
projections in Fig.~\ref{fig:tetra01} and Fig.~\ref{fig:raw2alltetra}. With those
plots, it becomes clear that the eigenplanes represent projections of a
simplex-like data into two-dimensions. Secondly, we note that the simplex
structure becomes less clear as one looks at planes corresponding to smaller
singular value directions; the signal eventually becomes buried in the noise. 

Similarly, the results of the factorization can be seen in eigenplanes from the
SVD of $E_{tf}W_{sf}=L_{tk}MN^T_{ks}$. These results (rows of $MN^T_{ks}$) are
shown in Fig.~\ref{fig:alltetra}, where we notice that the data is now
perfectly resides in simplex region as expected due to constraints.

\section{Coefficient of determination ($r^2$)}
\label{sec:pve}

We measured the goodness of the returns decomposition $R=EW$ by measuring the
coefficient of determination ($r^2$) as follows:

 \begin{equation}
	r^2 = 1 - SSE/SST
	\label{eq:r2}
\end{equation}

Here, SSE denotes the sum of square errors $||R - EW||_F^2$,
and SST is the total sum of squares $||R||^2_F$.  This is also known as the
\emph{proportion of variance explained} (PVE).  For the factorization of the
full dataset, normalized with the market mode removed, the calculated $r^2$ value is
$11.1\%$. The SVD of $R$ with singular values shown in Fig.~\ref{fig:rmt_comp} provides a convenient way to put this number in context for the returns dataset. Only 20 singular values (excluding the market mode) were above the cut-off that was predicted by random matrix theory for a matrix of purely random Gaussian entries. For any matrix $M$ with elements $m_{ij}$, the norm $||M||^2_F = \sum_{i,j} m_{ij}^2 = \sum_i s_i^2 $, where $s_i$ are the singular
values \citep{Press}. Thus, the fraction of intrinsic variation in $R$ above the cutoff is the sum of squares of the 20 singular values (not including market mode) divided by SST, $\sum_{i=1}^{i=20} s_i^2 / ||R||^2_F =19.8\%$. Therefore, as a first approximation, the factorization explains
$11.1/19.8=56\%$ of the \textit{random matrix theory (RMT) explainable variation}. \\

\begin{table}
\begin{center}
\begin{tabular}{|l|c|}
  \hline
  \textbf{Bulk Variation} & 80.2\% \\ \hline
  \textbf{Explainable Variation} & 19.8\% \\  \hline  
  \multicolumn{2}{ |c| }{} \\ \hline 
  \textbf{Factors} & \textbf{Percent of Explainable Variation} \\ \hline
  Market Mode (MM) & 8.0\% \\ \hline
  2 factors + MM & 26.0\% \\ \hline
  3  factors + MM & 36.1\% \\ \hline
  4  factors + MM & 42.8\% \\ \hline
  5  factors + MM & 48.9\% \\ \hline
  6  factors + MM & 55.3\% \\ \hline
  7  factors + MM & 59.4\% \\ \hline
  8  factors + MM & 63.7\% \\ \hline
  9  factors + MM & 68.1\% \\ \hline
 Fama and French & 24.0\% \\
  \hline
\end{tabular}
\end{center}
\caption{
\label{tab:PVE}
Percentage of the Explainable Variance captured by our model compared with the Fama and French factor model. Regression is done on the normalized dataset of 705 stocks without the market mode removed. To capture this, we add the market mode to factors obtained by our decomposition.}
\end{table}

\begin{figure*}[h] 
	\begin{centering} 
	\includegraphics[scale=.28]
	{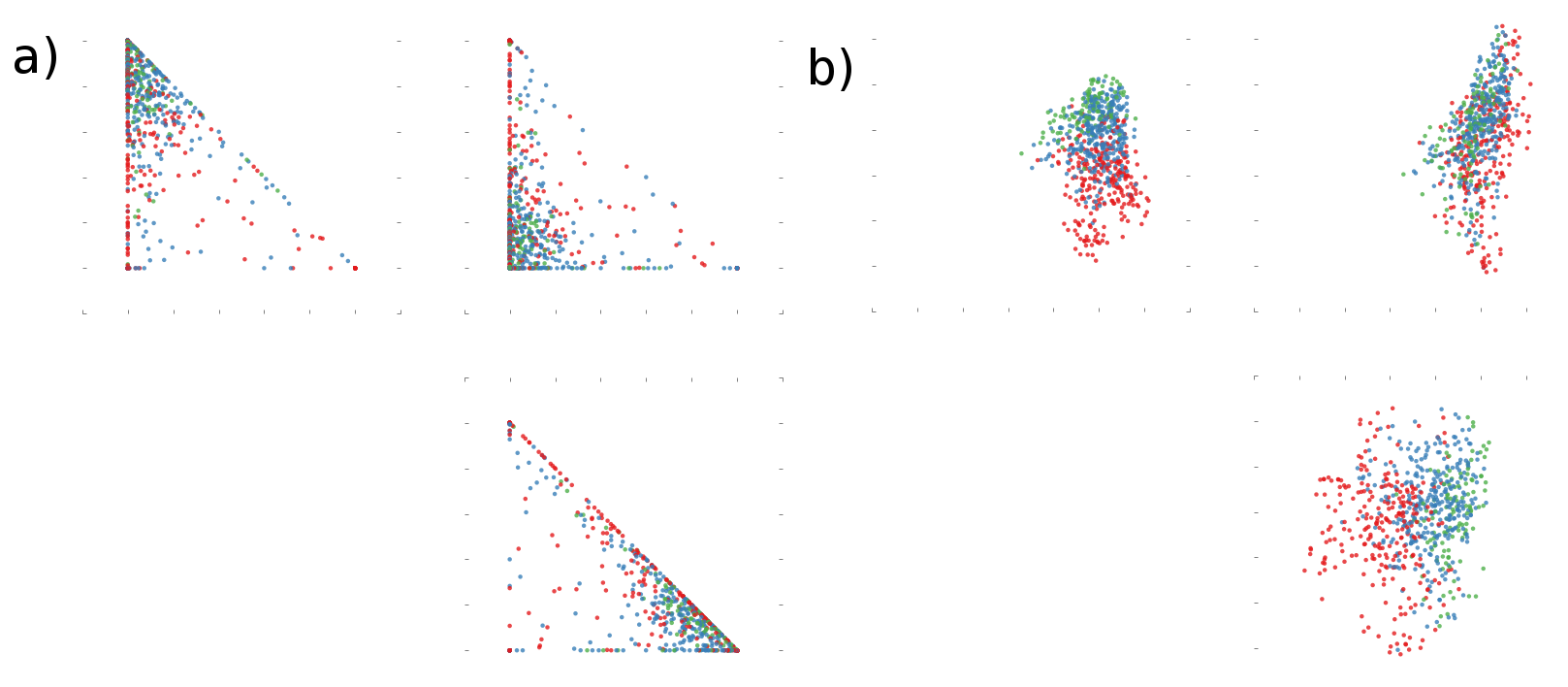} 		
	\caption{\label{fig:comp}\textbf{Three Factor Model vs.\ Fama and French.} 2D projections of the
	weights for each company in the SP500 with current tickers and data in the date range we 	
	consider. Red denotes companies with large market caps (market cap $ >$10 billion), blue 	
	denotes medium (market cap 2-10 billion) and green denotes small (market cap $<$ 2 	
	billion). For our decomposition (a),  there is no separation distinguishable by size of 
	company. In comparison, for the Fama and French decomposition (b), there appears a 
	gradation from large to small companies consistent with a factor of the model being related 
	to size. (This is natural, since one of Fama and French's factors explicitly is the difference 
	between large and small-cap returns). Thus our unsupervised 3-factor decomposition 
	appears quite distinct from Fama and French's hand-created one.} 
	\end{centering}
\end{figure*}

For reference we provide the RMT explainable variation for the factor decomposition of Fama and French, the classification by Scottrade\textsuperscript{\textregistered}, and the top 8 singular vectors given by SVD. The percentage of the RMT explainable variation for different numbers of factors compared to the 3 factor decomposition of Fama and French is shown in Table~\ref{tab:PVE}. Fama and French have the benefit of allowing factors to have positive or negative weights. In order to compare with another non-negative decomposition, we fix the weight matrix according to the Scottrade\textsuperscript{\textregistered} labels  and  run archetypal analysis for this $n=14$ factor version. The $r^2$ value for this decomposition is $10.7\%$ with a corresponding RMT explainable variance of $54.2\%$ compared to $56\%$ for our 8 factors. For completeness, we also note that if $R$ is rank-reduced to the eight stiffest components found by SVD (not including market mode), then the factorization explains $85\%$ of the the RMT explainable variation in $R$ with overall results in good accord with the analysis presented here. This implies that sector decomposition information was already contained in the stiff modes from the SVD of $R$, however SVD is not the appropriate tool for the decomposition. 
Fig.~\ref{fig:comp} further shows that our unsupervised 3-factor decomposition appears quite distinct from Fama and French's hand-created one.

\section{The Number $n$ of Canonical Sectors}
\label{sec:numbern}

It is an open problem to determine the effective dimensionality (optimal rank)
of a general dataset (matrix). One could select among models of different
dimensions using statistical tests such as the $r^2$ discussed above, or
information theory based criteria such as Akaike Information Criterion (AIC) or
the Bayesian Information Criterion (BIC), but the choice of the selection
criterion is itself generally made on an \textit{ad hoc} basis. Therefore, a
direct observation of the comprehensibility of results is often the most
reliable criterion.  In the dataset used for analysis described here, a
factorization with $n>8$ yielded results where both the emergent time series
$E_{tf}$ and weights in $W_{fs}$ showed qualitative signs of overfitting. For example,  with $n=9$ the results were in good agreement with $n=8$ except for an additional resulting sector involving participation from only 11 seemingly unrelated stocks (Table~\ref{tab:sec9} and Fig.~\ref{fig:sankey}). The high-level results of factorization with different values of $n$ may be explored in a number of ways, several of which are described below.

\begin{table}
\begin{center}
\begin{tabular}{|l l l|}
	\hline
	\textbf{Ticker} & \textbf{Company Name} &\textbf{Label}\\
	\hline
	EQT  &  EQT Corporation & Energy\\
	RDN  &  Radian Group Inc.  & Financials\\
	STT & State Street Corporation &Financials\\
	LH &  Laboratory Corp. of America Holdings & Healthcare \\
	UHS & Universal Health Services Inc.& Healthcare\\
	STZ & Constellation Brands Inc. & Non-Cyclicals\\
	CNL & Cleco Corporation & Utilities\\
	OKE & ONEOK Inc. & Utilities\\
	CAKE & The Cheesecake Factory Incorporated & Cyclicals \\
	EFX & Equifax Inc. & Industrials\\
	ESRX & Express Scripts Holding Company & Non-Cyclicals\\
	\hline
\end{tabular}
\end{center}
\caption{
\label{tab:sec9}
Companies which form a new sector when the dimensionality of the decomposition is increased from $n=8$ to $n=9$. The labels given are those indicated by Scottrade\textsuperscript{\textregistered}.}
\end{table}

\subsection{Sector Changes with Dimensionality}
\label{subsec:DimChanges}

One approach to investigating how the sector decomposition changes with dimension is to produce a flow diagram. To do this, we performed the fit $||E_{t,f}-E_{t,f'}S_{f',f}||_F^2$ with the constraint $\sum_{f'}S_{f',f}=1$. Hence the sectors for $n=9$ can be expressed as a linear combination of sectors for $n=8$, $n=8$ as a linear combination of $n=7$, and so forth. The results of these fits are presented in Fig.~\ref{fig:flows}.  The figure represents these relationships though connections between the decompositions for $n = N+1$ and $n=N$ weighted according to the matrix $S^{(N,N+1)}$.  More precisely, we create a node corresponding to each of the 9 sectors  whose size is proportional to $\sum_s W_{f,s}$ where $W_{f,s}$ is the weight matrix for the 9 sector decomposition. Hence, the relative node sizes represent the amount of the market particpating in the sector. Multiplying this vector by $S^{(8,9)}$ gives the approximate size for each node in $n=8$. Multiplying this vector by $S^{(7,8)}$ gives the approximate size for each node in $n=7$, and so on.  In this way, we generate a Sankey diagram whose node sizes correspond roughly to the amount of the market in the sector and whose connections depict how strongly the sectors for decompositions with different $n$ overlap. In the image, we see that the $n=9$ decomposition gives the 8 sector version with an additional small sector whose companies were listed in Table~\ref{tab:sec9}. We also see that for $n=7$ \textit{c-finance} and \textit{c-real estate} merge. At $n=6$, \textit{c-industrial} and \textit{c-cyclical} merge. For $n=5$, the new sector containing  \textit{c-industrial} and \textit{c-cyclical}  merges with \textit{c-non-cyclical}. For $n=4$, \textit{c-utility} and \textit{c-energy} merge. Finally, for $n=3$ and $n=2$, no clear pattern emerges given this image alone.

\subsection{Two and Three Sector Decompositions}

\begin{table*}[] \centering {\footnotesize 
\begin{tabular}{ |l l l l| l l l l|}
\hline
\textbf{c-assets} & \textbf{label} & \textbf{percent} & \textbf{full name} & \textbf{c-goods} & \textbf{label} & \textbf{percent} & \textbf{full name}  \\
\hline
DDR & real estate & 1.77\% & DDR Corp. & HON & tech & 0.53\% & Honeywell International Inc.\\
ONB & financial & 1.7\% & Old National Bankcorp.& TMO & health & 0.51\% & Thermo Fisher Scientific Inc. \\
BRE & real estate & 1.66\% & Brookfield Real Estate Serv. & NAV & cyclical & 0.49\% & Navistar International Corp. \\
PEI & real estate & 1.54\% & Pennsylvania RIT &CSL & basic & 0.47\% & Carlisle Companies Inc. \\
FMBI & financial & 1.5\% & First Midwest Bancorp. Inc. & IRF & tech & 0.47\% & International Rectifier Corp.\\
PRK & financial & 1.5\% & Park National Corp. & APD & basic & 0.46\% & Air Products \& Chemicals Inc.\\
BAC & financial & 1.42\% & Bank of America Corp. & PCP & basic & 0.43\% & Precision Castparts Corp.\\
STI & financial & 1.41\% & SunTrust Banks Inc. & OMC & misc services & 0.43\% & Omnicom Group Inc.\\
DRE & real estate & 1.29\% &Duke Realty Corp. & MXIM & tech & 0.43\% & Maxim Integrated Products, Inc.\\
UBSI & financial & 1.28\% & United Bankshares Inc. & TFX & health & 0.41\% & Teleflex Inc. \\
CPT & real estate & 1.28\% & Camden Property Trust &NSC & transport & 0.41\% & Norfolk Southern Corp. \\
PPS & real estate & 1.28\% & Post Properties Inc. & NBL & energy & 0.4\% & Noble Energy Inc.\\
WABC & financial & 1.26\% & Westamerica Bancorp. & SM & energy & 0.4\% & SM Energy Company\\
FMER & financial & 1.26\% & FirstMerit Corp. & WMT & retail & 0.39\% & Wal-Mart Stores Inc.\\
CNA & financial & 1.26\% & CNA Financial Corp. & CR & basic & 0.38\% & Crane Co.\\
VLY & financial & 1.25\% & Valley National Bancorp. & ADI & tech & 0.38\% &Analog Devices Inc.\\
MTB & financial & 1.24\% & M\&T Bankcorp. & ITW & cyclical & 0.38\% & Illinois Tool Works Inc.\\
WRI & real estate & 1.23\% & Weingarten Realty Investors & PPG & basic & 0.38\% & PPG Industries Inc.\\
BDN & real estate & 1.21\% & Brandywine Realty Trust & BA & capital & 0.38\% & The Boeing Company\\
ZION & financial & 1.2\% & Zions Bancorp. & AME & tech & 0.38\% & Ametek Inc.\\
\hline
\textbf{Total} & & 27.54\% & & \textbf{Total} & & 8.53\% & \\
\hline
\end{tabular}
}
\caption{
\label{tab:top20TwoSector}
Top 20 contributing companies to each sector in the two sector decomposition. Ranking is determined by the martix $C_{s,f}$ which describes each sector as a linear combination of stocks. Labels are those given by Scottrade\textsuperscript{\textregistered} and percentage describes the percentage of the sector attributable to the company.}
\end{table*}

\begin{table*}[] \centering {\footnotesize 
\begin{tabular}{|lll|lll|lll|}
\hline
\textbf{sector 1}&\textbf{label}&\textbf{percent}&\textbf{sector 2}&\textbf{label}&\textbf{percent}&\textbf{sector 3}&\textbf{label}&\textbf{percent} \\
\hline
XOM & energy & 1.29\% &BRE & real estate & 2.16\% &IRF & tech &1.29\% \\
HP & energy & 1.22\% &PEI & real estate & 2.08\% &EMC & tech &1.22\% \\
CVX & energy & 1.21\% &BWS & retail & 1.99\% &ADI & tech &1.21\% \\
ETR & utility & 1.2\% &CNA & financial & 1.79\% &CSCO & tech &1.2\% \\
APD & basic & 1.2\% &ONB & financial & 1.73\% &TXN & tech &1.2\% \\
OXY & energy & 1.19\% &DDR & real estate & 1.63\% &BMC & tech &1.19\% \\
NFG & utility & 1.18\% &PRK & financial & 1.59\% &SNPS & tech &1.18\% \\
PX & basic & 1.17\% &CBSH & financial & 1.59\% &PLXS & tech &1.17\% \\
CL & non-cyclical & 1.16\% &BC & cyclical & 1.56\% &CPWR & tech &1.16\% \\
NBL & energy & 1.15\% &FMER & financial & 1.55\% &AVT & tech &1.15\% \\
OII & energy & 1.11\% &RDN & financial & 1.54\% &SWKS & tech &1.11\% \\
LNT & utility & 1.11\% &MAS & capital & 1.54\% &HPQ & tech &1.11\% \\
D & utility & 1.08\% &DDS & retail & 1.47\% &PMCS & tech &1.08\% \\
DTE & utility & 1.07\% &FMBI & financial & 1.47\% &MXIM & tech &1.07\% \\
SCG & utility & 1.06\% &ALK & transport & 1.46\% &ARW & tech &1.06\% \\
WEC & utility & 1.04\% &WABC & financial & 1.43\% &TER & tech &1.04\% \\
APA & energy & 0.99\% &PCH & real estate & 1.42\% &ATML & tech &0.99\% \\
BAX & health & 0.98\% &VLY & financial & 1.41\% &MCHP & tech &0.98\% \\
MUR & energy & 0.98\% &BAC & financial & 1.41\% &LRCX & tech &0.98\% \\
CPB & non-cyclical & 0.98\% &STI & financial & 1.37\% &CGNX & tech &0.98\% \\
\hline
\textbf{Total} & & 22.38\% & \textbf{Total} & & 19.14\% & Total & & 32.18\% \\
\hline
\end{tabular}}
\caption{
\label{tab:top20ThreeSector}
Top 20 contributing companies to each sector in the three sector decomposition. Ranking is determined by the martix $C_{s,f}$ which describes each sector as a linear combination of stocks. Labels are those given by Scottrade\textsuperscript{\textregistered} and percentage describes the percentage of the sector attributable to the company.}
\end{table*}

\begin{table*}[] \centering {\footnotesize 
\begin{tabular}{|ll|ll|ll|}
\hline
\textbf{sector 1}&\textbf{full name}&\textbf{sector 2}&\textbf{full name}&\textbf{sector 3}&\textbf{full name} \\
\hline
XOM & Exxon Mobil Corp. & BRE & Brookfield Real Estate Serv. & IRF & International Rectifier Corp. \\
HP & Helmerich \& Payne Inc. & PEI & Pennsylvania RIT & EMC & EMC Corp. \\
CVX & Chevron Corp.  & BWS & Brown Shoe Co. Inc. & ADI &  Analog Devices Inc.\\
ETR & Entergy Corp. & CNA & CNA Financial Corp. & CSCO & Cisco Systems Inc. \\
APD & Air Products \& Chemicals Inc. & ONB & Old National Bancorp. & TXN & Texas Instruments Inc. \\
OXY & Occidental Petroleum & DDR & DDR Corp. & BMC & BMC Software Inc. \\
NFG & National Fuel Gas Company & PRK & Park National Corp. & SNPS & Synopsys Inc. \\
PX & Praxair Inc. & CBSH & Commerce Bancshares Inc. & PLXS & Plexus Corp. \\
CL & Colgate-Palmolive Co. & BC & Brunswick Corp. & CPWR & Compuware Corp. \\
NBL & Noble Energy Inc. & FMER & FirstMerit Corp. & AVT & Avnet Inc.\\
OII & Oceaneering International Inc. & RDN & Radian Group Inc. & SWKS & Skyworks Solutions Inc. \\
LNT & Alliant ENergy Corp. & MAS & Masco Corp. & HPQ & Hewlett-Packard Company\\
D &  Dominion Resources Inc. & DDS & Dillard's Inc. & PMCS & PMC-Sierra Inc. \\
DTE & DTE Energy Corp. & FMBI & First Midwest Bancorp. Inc. & MXIM & Maxim Integrated Products Inc.\\
SCG & SCANA Corp. & ALK & Alaska Air Group Inc. & ARW & Arrow Electronics Inc.\\
WEC & Wisconsin Energy Corp. & WABC & Westamerica Bancorp. & TER & Teradyne Inc. \\
APA & Apache Corp. & PCH & Potlatch Corp. & ATML & Atmel Corp. \\
BAX & Baxter International Inc. & VLY & Valley National Bancorp. & MCHP & Microchip Technology Inc. \\
MUR & Murphy Oil Corp. & BAC & Bank of America Corp. & LRCX & Lam Research Corp.\\
CPB & Campbell Soup Company & STI & SunTrust Banks Inc. & CGNX & Cognex Corp. \\
\hline
\end{tabular}}
\caption{
\label{tab:top20ThreeSectorFullNames}
Top 20 contributing companies to each sector in the three sector decomposition. Ranking is determined by the martix $C_{s,f}$ which describes each sector as a linear combination of stocks.}
\end{table*}

\begin{figure*}[h]
	\begin{centering}
	\includegraphics[scale=0.3]
	{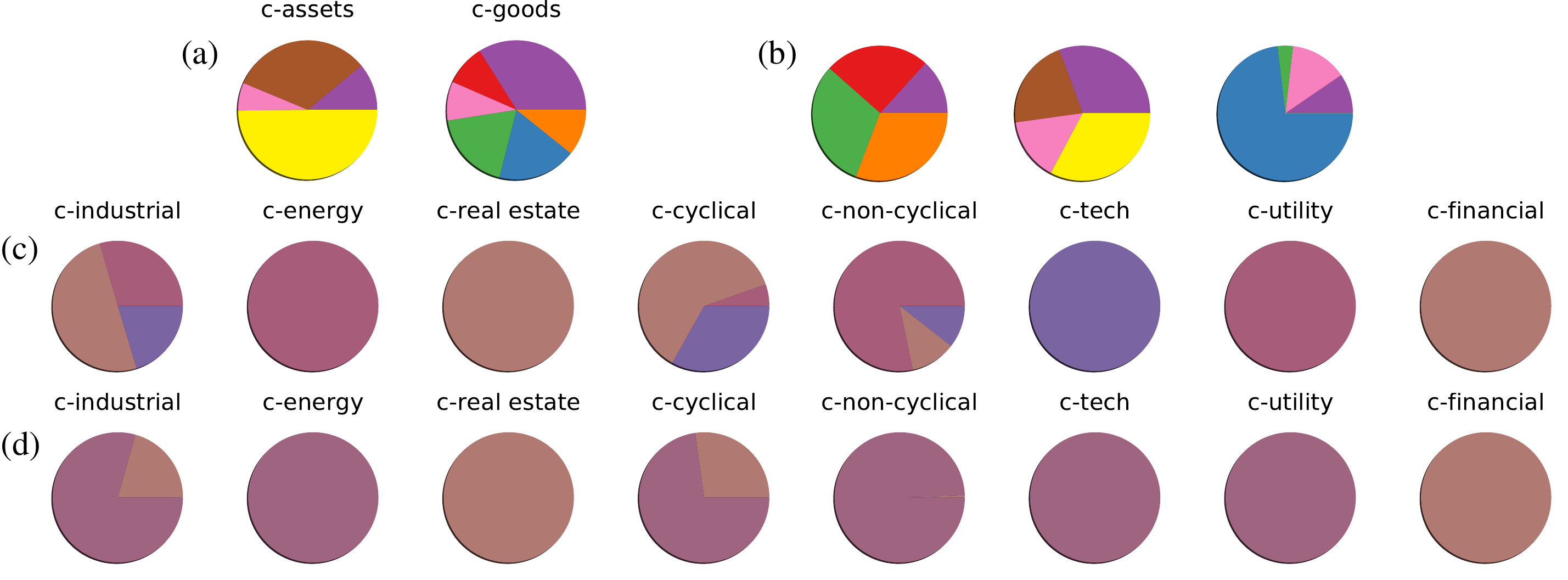}
	\caption{\label{fig:pies2} \textbf{Pie charts depicting sectors as linear combinations of other sector decompositions having a different value of the dimensionality $n$}. (a) Two sector decomposition with respect to the eight sector version (b) Three with respect to eight (c) eight with respect to two (d) eight with respect to three.  For (a) and (b) the color scheme is the same as used throughout for the eight sector decomposition. For (c) and (d) colors correspond to those in Fig.~\ref{fig:sankey} for the two and three sector nodes. Through these charts it is evident that the two sector decompositions corresponds to an \textit{c-assets} sector containing \textit{c-finance} and \textit{c-real estate}. and a \textit{c-goods} sector containing companies which provide goods and services. In (c) and (d) we see \textit{c-industrial}, \textit{c-cyclical} and \textit{c-non-cyclical} which merge by $n=5$ split between the two and three factor decompositions respectively, consistent with Fig.~\ref{fig:sankey}.  }
	\end{centering}
\end{figure*}

We further explore the two and three sector decompositions by examining their constituent companies and looking at pie charts describing the relationship between our 8 sector decomposition and those with $n=2$ and $n=3$ respectively. Recall that each archetype is constrained to be a linear combination of companies, or in other words to lie in the convex hull of the data. Using this information,  we list the 20 companies which contribute the most to each sector in the two and three factor decompositions (Tables~\ref{tab:top20TwoSector}, \ref{tab:top20ThreeSector} and \ref{tab:top20ThreeSectorFullNames}). For the two sector decomposition, we find the sectors divide roughly into \textit{c-assets} (e.g. financial and real estate companies) and \textit{c-goods} (e.g. companies which provide goods and services). For $n=3$, the division is less clear. Another way to look at the constituents of these sectors is by examining pie chart representations of these decompositions. Again consider the fit $||E_{t,f}-E_{t,f'}S_{f',f}||_F^2$ with the constraint $\sum_{f'}S_{f',f}=1$. Applying this, we can express the two sector archetypes as linear combinations of the 8 sector archetypes and vice versa. Additionally, we can do the same for the three factor decomposition. The pie charts these fits produce are shown in Fig.~\ref{fig:pies2}. The results are consistent with the sector breakdowns described from examining the constituent companies. 

\subsection{Robustness}

In general, a factorization analysis of the returns dataset would be sensitive
to number of stocks in the dataset, criteria applied for picking stocks, period
over which historical prices are obtained, and frequency at which returns are
computed. A robust macroeconomic analysis would therefore require a large
number of stocks chosen without sampling bias, with returns calculated over the
period of interest and sensitivity checked for frequency of returns
calculation. On the other hand, an equity fund manager faces a less daunting
task for an analysis that is limited to the universe of her portfolio of
stocks: either to find its canonical sectors, or to analyse the exposure of
her holdings to the core sectors of the economy.

\section{Canonical Sector Indices}\label{sec:indices}

The matrix $C_{sf}$ in decomposition $R=RCW$ represents how returns $R$ of
stocks $s$ must be combined to make canonical sector returns $E_{tf} =
R_{ts}C_{sf}$. Since a canonical sector is defined as a combination of stocks,
an investment in the sector $f$ can made via buying a basket of constituent
stocks $s$ in proportions given by $C_{sf}$ or through an index $I_{tf}$:
\begin{equation}
	 I_{tf} = p_{ts'}C_{s'f}
	\label{eq:index}
 \end{equation}
where, $p$ are stocks prices suitably weighted
by market cap or other divisor as common practice for common
indices \citep{Tagiliani}. An unweighted index of this kind is shown in the
bottom row of Fig.~\ref{fig:retpanels} for results corresponding to the
analysis described in this paper. Conversely, a pre-defined basket of stocks
such as the S\&P 500\textsuperscript{\textregistered} can be unbundled to find its exposure to the canonical
sectors. With an investment strategy employing longs and shorts at the same
time in correct proportions, it is conceivable to invest in, for example, the
\textit{c-tech} component of S\&P 500\textsuperscript{\textregistered}.

The desirable features of an index include completeness, objectivity and
investability \citep{Pastor13}. The \textit{c-indices} constructed using the
ideas outlined here would not only be of value to investors through investment
vehicles such as ETFs, Futures, etc., but also serve as important economic
indicators. 

\bibliographystyle{rQUF}
\bibliography{HaydBib}

\end{document}